\documentclass[aps,showpacs,floatfix,twocolumn,prd]{revtex4}
\usepackage{epsfig,graphics,color}
\usepackage{graphicx}
\usepackage{dcolumn}
\usepackage{bm}
\usepackage{amsmath,amssymb}
\usepackage{eucal,eufrak}

\newcommand \tie {{\it i.e.}}

\newcommand \eg {{\it e.g.}}  
\newcommand \f {\not\!}

\newcommand \p {{\prime}}

\newcommand \ra  {\rightarrow}

\newcommand \vk {\vec{k}}
\newcommand \vl {\vec{l}}

\newcommand \vp {\vec{p}}

\newcommand \A {\alpha}
\newcommand \B {\beta}
\newcommand \g {\gamma}

\newcommand \D {\Delta}
\newcommand \kd  {\delta}

\newcommand \e {\epsilon}

\newcommand \h {\theta}
\newcommand \ro {\rho}

\newcommand \ld {\lambda}

\newcommand \md {\mathfrak{D}}

\newcommand \Op {{\mathcal O}}

\newcommand \x {\cdot}
\newcommand \hf {\frac{1}{2}}

\newcommand \lc {\langle}
\newcommand \rc {\rangle}
\newcommand {\lb} {\left[}
\newcommand {\rb} {\right]}
\newcommand {\lp} { \left( }
\newcommand {\rp} {\right)}
\newcommand \lt {\left}
\newcommand \rt {\right}
\newcommand \prt {\partial}
\newcommand \nt {\noindent}

\newcommand \gmn {g^{\mu \nu}}

\newcommand \bvec{\left( \begin{array}{c} }
\newcommand \evec{\end{array} \right)}
\newcommand \tr {\mbox{{\bf Tr}}}
\newcommand \bea{\begin{eqnarray} }
\newcommand \eea{\end{eqnarray} }
\newcommand \nn {\nonumber}
\newcommand \be {\begin{equation}}
\newcommand \ee {\end{equation}}

\newcommand \mbx {\mbox{}}

\newcommand \psibar {\bar{\psi}}
\newcommand \ata {& \times &}
\newcommand {\apa} {& + &}
\newcommand {\aqa} {&=&}

\newcommand \slm {\sum\limits}

\begin{document}

\title{Hard collinear gluon radiation and multiple scattering in a medium}

\author{Abhijit~Majumder}
\affiliation{Department of Physics, The Ohio State University, Columbus, Ohio 43210.}

\date{\today}

\begin{abstract} 
The energy loss of hard jets produced in the Deep-Inelastic scattering (DIS) off a large nucleus is considered 
in the collinear limit. In particular, 
the single gluon emission cross section due to multiple scattering in the medium is calculated.
Calculations are carried out in the higher-twist scheme, which 
is extended to include contributions from multiple transverse scatterings on both the produced quark and the radiated gluon. 
The leading length enhanced parts of these power suppressed contributions are resummed. 
Various interferences between such diagrams lead to the Landau-Pomeranchuk-Migdal (LPM) effect. 
We resum the corrections from an arbitrary number of scatterings and isolate the leading contributions
which are suppressed by one extra power of the hard scale $Q^{2}$. All powers of the emitted 
gluon forward momentum fraction $y$ are retained.
We compare our results with the previous calculation of single scattering per emission in the higher-twist 
scheme as well as with multiple scattering resummations in other schemes. 
It is found that the leading ($1/Q^2$) contribution to the double differential gluon production cross section, 
in this approach, is equivalent to that obtained from the single scattering 
calculation once the transverse momentum of the final quark is integrated out. 
We comment on the generalization of this formalism to Monte-Carlo routines.
\end{abstract}


\maketitle


 \section{introduction}


The modification of hard jets in dense media~\cite{quenching,Baier:1996kr,Zakharov:1996fv}, 
and the use of this modification to measure the 
partonic structure of both confined or deconfined matter is now a considerably mature science~\cite{white_papers}. 
While the modification of full jet structure is only now being measured, the modification of the yield 
of leading hadrons, (a reduction or suppression of the yield) has been measured by the HERMES 
experiment at DESY for Deep-Inelastic Scattering (DIS) on 
a large nucleus~\cite{Airapetian:2000ks} and by both the STAR and PHENIX detectors~\cite{highpt} at 
the Relativistic Heavy Ion Collider (RHIC) for jets in heavy-ion collisions. 
On the theory side, there now exist at least four sophisticated jet modification 
schemes which describe the single particle 
yields within a pQCD based phenomenology. It has now become customary to refer to these as the 
Gyulassy-Levai-Vitev (GLV) scheme~\cite{GLV}, the Arnold-Moore-Yaffe scheme (AMY)~\cite{AMY}, 
the Armesto-Salgado-Wiedemann scheme~\cite{ASW} and the higher-twist (HT) 
approach~\cite{HT,Zhang:2003yn,Majumder:2007hx,Majumder:2007ne}. 
Of these, the GLV, ASW and HT approaches have also been applied to the case of DIS 
on a large nucleus~\cite{Vitev:2007ve,Arleo:2002kh,Wang:2002ri} using almost an identical formalism 
to that used in the case of hot matter. All these calculations obtained a much smaller value of the 
leading transport coefficient in cold nuclear matter 
than for heavy-ion collisions. This represents 
one of the leading pieces of evidence for the formation of a new type of matter in high energy heavy-ion collisions.

The transport coefficient in question, $\hat{q}$, defined as the transverse 
momentum squared gained by a hard parton per unit length, has been identified 
as the leading effect of the medium on a hard parton~\cite{Baier:1996kr,Majumder:2007hx} for the 
case of light flavor suppression. In the case of DIS on a nucleus, the $\hat{q}$, assumed 
to scale with the nucleon density, 
is a constant, as a function of time, over most of the space through which the parton 
traverses (ignoring edge effects in the Woods-Saxon distribution). The case of 
heavy-ion collisions requires more sophistication: $\hat{q}$ is usually assumed to scale with 
an intrinsic quantity such as the entropy density $ s \propto T^3$ ($T$ is the local temperature). In a 
dynamically expanding medium, $T$ depends on both the location in space and time. 
Hence a complete calculation requires an underlying realistic 
model of the medium. Three of the formalisms (ASW, AMY and HT) have been extended to 
calculations where the evolution of the medium is treated with a 3-D relativistic fluid dynamical 
simulation~\cite{Bass:2008rv}.

Current experiments have, however, progressed far beyond a measurement of light flavor single particle 
inclusive suppression and now include measurements of heavy-flavor suppression and azimuthal 
anisotropy~\cite{Abelev:2006db,Adare:2006nq} as well as triggered distributions of associated 
particles. Theory calculations have also progressed, especially in the case of  two particle correlations. 
While the GLV~\cite{Vitev:2007ve,Qiu:2003pm}, ASW~\cite{Renk:2006pk} and AMY~\cite{Qin:2009bk} 
approaches have attempted to describe the away side associated yield using a leading order partonic cross section, 
the HT has incorporated a Next-to-Leading Order (NLO) hard cross section~\cite{Zhang:2007ja}.  
Along with this, the HT approach has also remained the sole formalism to incorporate multi-hadron 
fragmentation functions in an effort to explain the near side associated yield~\cite{maj04d}.

In spite of these developments, the HT approach still suffers from the disadvantage that, unlike the ASW and the AMY 
schemes, it is a single scattering scheme. In its original form~\cite{HT}, the HT approach only contained one
scattering and one emission in the medium. In subsequent efforts~\cite{Majumder:2009zu}, the HT 
approach was extended to include multiple emissions by iterating the single-scattering-single-emission kernel 
by means of a Dokshitzer-Gribov-Lipatov-Altarelli-Parisi (DGLAP)~\cite{gri72} evolution equation. 
While this does include the case where a hard parton or its radiated gluons scatter multiple times, these 
scatterings are included incoherently, i.e., there is at most one scattering within the formation time of a single produced gluon. 
The applicability of this assumption can be quickly estimated: imagine a typical high energy DIS event (say at the HERMES experiment~\cite{Airapetian:2007vu}) with a $Q^2 \sim 10$GeV$^2$ and a forward energy $\nu \sim 30$GeV. The produced quark is off-shell by at most $Q^2$ and this depreciates to a non-perturbatively small value in a series of iterative steps. Consider a step where the drop is $\D Q \sim 2$GeV with a momentum fraction of 1/2,  
the formation length of this radiation is, 
\bea
\tau_f &\sim& \frac{ \nu}{   2(\D Q)^{2}  } \sim 1 {\rm fm}.
\eea
Assuming that there is at most one scattering per nucleon, given that double scattering in a nucleon involves 
twist-4 matrix elements, and there is approximately one nucleon per fm of length, this particular emission will 
encounter one scattering during the formation of the radiated gluon. However, for smaller drops in $Q^{2}$, smaller 
initial values of $Q^{2}$ and higher energies $\nu$, this simple approximation breaks down. The limit of at 
most one scattering per radiation cannot hold in the case of jets at RHIC either, as the density of scatterers is much 
larger than in cold nuclear matter. Such considerations require the extension of the HT scheme to include 
multiple coherent scatterings per emitted gluon. It is the object of this paper to carry out this extension.

This paper completes the efforts begun in Refs.~\cite{Majumder:2007hx,Majumder:2007ne} 
by computing the single gluon emission cross section from multiple scattering of a hard quark in a dense medium. 
The prior efforts 
included the simpler problems of evaluating the all twist contribution to the propagation of a hard quark produced 
in DIS on a large nucleus without gluon radiation and also the emission of electromagnetic 
radiation from this hard quark. 
Similar to the case of photon radiation at all twist in 
Ref.~\cite{Majumder:2007ne}, we will evaluate the triple differential distribution to split the 
hard quark into a quark and a gluon separated by a transverse momentum $\vl_{q \perp}$ with the gluon carrying a large 
transverse momentum $l_{\perp}$ and a fraction $y$ of the original quark light-cone momentum. 
In a real experiment, the size of the electromagnetic fine structure constant $\A_{EM}$ ensures that 
multiple hard photon bremsstrahlung is suppressed, however given the size of $\A_s$ at the typical 
scales involved, the produced quark will radiate gluons multiple times. 
In this effort, only the differential single gluon emission spectrum will 
be calculated. The inclusion of multiple emissions which involves an iteration of the single gluon emission kernel, 
followed by the eventual numerical calculation and 
comparison with experimental data will be presented in a separate effort.

The paper is organized as follows: Sec.~II will introduce the basic notation by evaluating the leading 
twist contribution to the differential collinear gluon radiation spectrum from a hard quark produced in DIS on a large nucleus. 
Sec.~III will identify an inclusive class of diagrams that yield leading contributions to the all-twist gluon emission 
spectrum. The collinear approximation will be invoked, integrations over the irrelevant momenta will be 
carried out followed by simplifications in the numerator structure. Sec.~IV 
will carry out the various pole integrations and identify the source of the Landau-Pomeranchuk-Migdal effect. 
Sec.~V will include discussions of the color factors and length enhancements of different classes of all-twist diagrams. 
Sec.~VI will evaluate the all-twist contributions in two different limits and compare with the calculations in the 
ASW and AMY formalisms. In Sec.~VII we present our concluding discussions.


 \section{Leading  twist  and collinear gluon radiation}


In this section we calculate the next-to-leading order
correction to semi-inclusive 
DIS on a large nucleus with a quark in the final state. By next-to-leading order, we simply mean 
including one interaction term in the amplitude and complex conjugate, which converts a  
single quark in to a quark and a gluon. 
As in the previous 
efforts ~\cite{Majumder:2007hx,Majumder:2007ne} we will assume and refrain from an attempt to 
discuss factorization~\cite{Qiu:1990xx,Qiu:1990xy}. 
 We re-derive this straightforward  
result to familiarize the reader with our conventions and the choice 
of gauge. 

\begin{figure}[htbp]
\resizebox{3in}{1.5in}{\includegraphics[0in,0in][8in,4in]{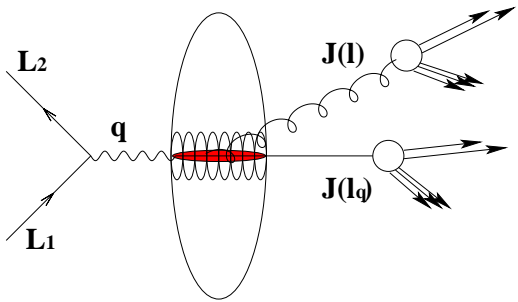}} 
    \caption{ The Lorentz frame chosen for the process where a nucleon in a large nucleus is
    struck by a hard space-like photon leading to the production of an outgoing parton and a radiated gluon.}
    \label{fig1}
\end{figure}

Consider the semi-inclusive 
process of DIS off a nucleus in the Breit frame where  one quark with a transverse 
momentum $l_{q_\perp}$ and a  bremsstrahlung gluon with transverse 
momentum $l_\perp$ are produced (see Fig.~\ref{fig1}),  
\bea
{\mathcal L} (L_1) + A(p) \longrightarrow  \mathcal{L} (L_2) + q( l_{q_\perp} ) + G(l_\perp) + X .
\label{chemical_eqn}
\eea
\nt
In the above equation, $L_1$ and $L_2$ represent the momentum of the 
incoming and outgoing leptons. The incoming nucleus of atomic mass 
$A$ is endowed with a momentum $Ap$. In the final state,
all high momentum hadrons ($h_1,h_2,...$) with momenta $p_1,p_2,\ldots$ are detected and their
momenta summed to obtain the quark jet momentum $(l_q)$ and gluon jet momentum $(l)$ and $X$ denotes that the 
process is semi-inclusive.

Our choice of light-cone component notation for four vectors ($p \equiv [p^+,p^-, \vec{p}_\perp]$)  
is somewhat different from the regular notation, i.e., 
\bea
p^+ = \frac{p^0 + p^3}{2}; \,\,\, p^- = p^0 - p^3.
\eea
In spite of the asymmetric definition, the formal structure of all dot products will be identical to 
the standard notation. 
The kinematics is defined in the Breit frame as sketched in Fig.~\ref{fig1}. 
In such a frame, the incoming virtual photon $\g^*$ and the nucleus have 
momentum four vectors $q,P_A$ given as, 

\[
q = L_2 - L_1 \equiv \left[\frac{-Q^2}{2q^-}, q^-, 0, 0\right], 
\mbox{\hspace{1cm}}
P_A \equiv A[p^+,0,0,0].
\]

\nt
In this frame, the Bjorken momentum fraction of the incoming quark with respect to its parent nucleon 
which has the average momentum $[p^{+},0,0,0]$ is given as 
$x_B = Q^2/2p^+q^-$. After the scattering with the virtual photon, the final state quark has a 
four momentum $\sim [0,q^{-},0,0]$ i.e., a large momentum in the negative $z$ or negative light cone 
direction. This outgoing quark will radiate a gluon. The radiated gluon has a transverse momentum of 
$l_\perp$ and carries a fraction $y$ of the forward momentum $q^-$ of the 
quark originating in the hard scattering, \tie, 

\bea
y = \frac{l^-}{q^-} .
\eea

The double differential cross section of the semi inclusive process 
with a final state quark (which will eventually become a jet) with transverse momentum $l_{q_\perp}$  and a final 
state gluon with transverse momentum $l_\perp \gg \Lambda_{QCD}$ may be expressed as 

\bea
\frac{E_{L_2} d \sigma } {d^3 L_2 d^2 l_{q_\perp}  d^2 l_\perp dy  } &=&
\frac{\A_{em} ^2}{2\pi s  Q^4}  L_{\mu \nu}  
\frac{d W^{\mu \nu}}{d^2 l_{q_\perp} d^2 l_\perp dy}, \label{LO_cross}
\eea

\nt
where $s = (p+L_1)^2$ is the total invariant mass of the lepton nucleon 
system. In the single photon exchange approximation, the leptonic part of the cross section is described 
by the particular form of the leptonic tensor denoted as $L^{\mu \nu}$ given as,
\bea 
L_{\mu \nu} = \frac{1}{2} \tr [ \f L_1 \g_{\mu} \f L_2 \g_{\nu}].
\eea

\nt
In the notation used in this paper, $| A; p \rc$ represents the spin averaged initial state of 
an incoming nucleus with $A$ nucleons with a momentum $p$ per nucleon. The 
general final hadronic or partonic state is defined as 
$| X \rc $.
As a result, the semi-inclusive hadronic tensor in the nuclear state $|A\rc$ may be defined as 
\bea 
{W^A}^{\mu \nu}\!\!\!\!&=& \!\!\!\! \sum_X  (2\pi)^4 
\kd^4 (q\!+\!P_A\!-\!p_X ) \nn \\
\ata \lc A; p |  J^{\mu}(0) | X  \rc \lc X  | J^{\nu}(0) | A;p \rc  \nn \\
&=& 2 \mbox{Im} \left[  \int d^4 y_0 e^{i q \cdot y_0 } \lc A;p | J^{\mu} (y_0) J^{\nu}(0) | A;p \rc \right].
\eea
In the equation above, 
the sum ($\sum_X$) runs over all possible hadronic states and 
$J^{\mu} =  Q_q \bar{\psi}_q \g^\mu \psi_q$ is the 
hadronic electromagnetic current, where, $Q_q$ is the 
charge of a quark of flavor $q$ in units of the positron charge $e$. 
It is understood that the quark operators are written in the  interaction picture and two 
factors of the electromagnetic coupling constant $\A_{em}$ have already been extracted and 
included in Eq.~\eqref{LO_cross}.
The leptonic tensor will not be discussed further. The focus in the remaining 
shall lie exclusively on the hadronic tensor. This tensor will be expanded 
order by order in a partonic basis with one on-shell gluon in the final state 
and leading twist and maximally length enhanced higher 
twist contributions will be isolated.

The leading twist contribution is obtained by expanding the products of 
currents to next-to-leading order in $\A_s$ to account for the radiated gluon. 
This contribution may be expressed diagrammatically 
as in Fig.~\ref{fig1}. This represents the process where a hard quark, 
produced from one nucleon in a deep-inelastic scattering on a nucleus, 
proceeds to radiate a hard gluon and then exits the nucleus without further interaction. 
Diagrams where the hard gluon is emitted from the incoming quarks are suppressed in our
choice of light cone gauge. In the following, we analyze this contribution in 
some detail. Indeed, there is no new information presented in the current 
section and the discussion of such contributions is now well established~\cite{gri72}. 
The approximations made in this section will form the 
basis for the analysis of single inclusive gluon radiation at all twist. The semi-inclusive
hadronic tensor may be expressed as, 
\bea
{W_0^A}^{\mu \nu} 
%
&=&  A C_p^A W_0^{\mu \nu} \label{w_mu_nu_twist=2}\\
&=& A C_p^A \int d^4y_0 \lc p |  \psibar(y_0) \g^\mu  \widehat{\Op^{00}} \g^\nu \psi(0) | p \rc  \nn \\
&=& C_p^A \int d^4 y_0\tr [\frac{\g^-}{2} \g^\mu \frac{\g^+}{2} \g^\nu ] F(y_0) \Op^{00}(y_0) .\nn 
\eea
In the above equation, $C_p^A$ expresses the probability to 
find a nucleon state with momentum $p$ inside a nucleus with $A$ nucleons. 
In the collinear 
limit, the incoming parton is assumed to be endowed with very high forward momentum $({p_0}^+ = x_0 p^+, p_0^- \ra 0)$
with negligible transverse momentum ${p_0}_\perp \ll p_0^+$.  Within the 
kinematics chosen, the incoming virtual photon also has no transverse 
momentum. As a result, the produced final state parton also has a vanishingly small transverse momentum 
(\tie, with a distribution $\kd^2(\vec{p}_\perp)$). 
In this limit, the leading spin projection of  the pieces which represent the initial state 
and final state may be taken. The factors, 
\bea
\g^+ = \frac{\g^0 + \g^3}{2} \,\,\, ;\,\,\,\,\,    \g^- = \g^0 - \g^3, 
\eea
are used to obtain  the spin projections along the leading momenta of the outgoing state and the incoming 
state. The coefficients of these projections are the two functions, 
\bea
F(y_0) = A \lc p | \psibar(y_0) \frac{\g^+}{2} \psi(0) | p \rc \label{F(y_0)}
\eea
\nt and (in a notation where the superscript on the operator $\Op^{00}$ implies that the 
quark undergoes no scattering in the initial or final state)

\bea 
\Op^{00} &=& \tr \left[ \frac{\g^-}{2} \widehat{\Op^{00}} \right]  \label{O_00_1}\\
 &= & \int \frac{d^4 l}{(2\pi)^4}   d^4 z d^4 z^\p 
\frac{d^4 l_q}{(2\pi)^4}\frac{d^4 p_0}{(2\pi)^4} \frac{d^4 p^\p_0}{(2\pi)^4}\nn \\ 
\ata   \tr \left[  \frac{\g^-}{2} \frac{-i (\f p_0 + \f q )}{ (p_0 + q )^2 - i \e }  
i\g^\A  \f l_q   2\pi \kd (l_q^2)  \right.  \nn \\
\ata \left. G_{\A \B} (l)   2\pi \kd (l^2)  (-i\g^{\B})
\frac{ i (\f p_0^\p + \f q )}{ (p_0^\p + q )^2 +  i \e }  \right] \nn \\
\ata e^{i q \x y_0 }e^{ -i (p_0 + q) \x ( y_0 - z) } 
e^{-i l \x (z - z^\p)} e^{-i l_q \x (z - z^\p)} \nn \\
\ata e^{ -i (p_0^\p + q) \x z^\p  } g^2  . \nn
\eea

\nt
Integrating over $z$ and $z^\p$ yields the two four-dimensional $\kd$-functions: 
$\kd^4 ( p_0 + q - l - l_q )$ and $\kd^4 (p_0 - p_0^\p )$.

The reader will have noted that we have ignored various projections such as those which 
arise from the ($\perp$)-components of the $\gamma$ matrices, \eg,
\bea
C_P^A \tr \left[   \frac{\g^{\perp_i}}{2} \g^\mu \frac{\g^{\perp_j}}{2} \g^{\nu} \right] F_{\perp_i} (y_0) \Op^{00}_{\perp_j}.
\eea
\nt
This approximation may be justified in the high energy, collinear limit $l_\perp^2/y << Q^2$ where such contributions are 
suppressed compared to those of Eq.~\eqref{w_mu_nu_twist=2}. As pointed out, in this effort, we will often adjudicate the 
importance of different terms using scaling arguments inspired by SCET. The kinematic regime explored in this 
article is $l_\perp \sim \lambda Q$ and both $y, (1 - y) \lesssim 1$. In this region, $l_\perp^2/y \sim \ld^2 Q^2 \ll Q^2$. Another 
interesting regime is the soft $y$ limit where $y \sim \ld$; even in this region one may ignore the 
non-transverse projections of the virtual photon.
In so doing, the focus of the remainder of 
this article has been limited to projections where the incoming virtual photon is transverse.

The on-shell $\delta$-function over $l$ is used to set $l^+ = l_\perp^2/2l^-$. 
The other on-shell $\kd$-function, instills the condition, 
\bea
\kd(l_q^2) &=& \kd[ ( p_0 + q - l )^2 ]  \nn \\ 
&\simeq& \kd [ -Q^2 + 2p_0^+ (q^- - l^-) - 2q^+ l^- - 2q^- l^+  ]     \nn \\
\aqa \frac{1}{2 p^+ q^- } \kd \left[ x_0 (1 - y) - x_B ( 1 - y  )  - \frac{l_\perp^2}{2 p^+ q^- y } \right] \nn \\
\aqa \frac{\kd [x_0 - x_B - x_L ]}{2 p^+ q^- (1- y)} , \label{delta_l_q}
\eea

\nt
where, the collinear condition that $p_{0_\perp} \ra 0$ has been used  to simplify the final 
equation. The new \emph{momentum fraction} $x_L$ has been introduced: 

\bea
x_L = \frac{l_\perp^2}{ 2 p^+ q^- y (1-y) } = \frac{1}{p^+ \tau_f} \sim \ld^2, 
\eea
where 
$y$ has already been defined as the momentum fraction of the radiated photon ($ l^-/q^-$)
and $\tau_f \sim 1/(\ld^2 Q)$ is the formation time of the radiated gluon.

The factor $G_{\A \B}(l)$ in Eq.~\eqref{O_00_1} represents the radiated 
photon's spin sum.  In this effort, the light cone gauge ($A^-=0$) will be assumed, 
\tie,
 
\bea
G_{\A \B} (l) &=& -\gmn + \frac{l^\mu n^\nu + l^\nu n^\mu}{ n\x l} , \label{light_cone_polarization}
\eea

\nt
where we have introduced the light cone vector $n \equiv [1,0,0,0]$ which yields 
$l \x n = l^-$. Note that with this choice of gauge, the largest component of the 
vector potentials from the initial states may still be regarded as the ($+$)-components. 

Substituting the above simplifications in  Eq.~\eqref{O_00_1}, 
leads to the simplified form for the final state projection: 

\bea
\Op^{00} &=& \int \frac{d^4 p_0}{(2 \pi)^4} 
\frac{d l^- d^2 l_\perp}{(2 \pi)^3 2l^-} g^2 e^{ -i p_0 \x y_0} \nn \\
\ata \tr \left[ \frac{\g^-}{2} 
\frac{\g^+ q^-  }{2p^+q^-( x_0 - x_B - {x_D}_0  - i \e)} \right.  \nn \\
\ata  
\left\{  \g_\perp^\A \g^- ( [x_0 - x_B] p^+  - l^+   )  
\g_\perp^\B (-{g_\perp}_{\A \B})  \right. \nn \\
&-&  \frac{ \g_\perp \x l }{l^-} \g_\perp \x  l    \g^- 
- \g^- \g_\perp \x  l  \frac{ \g_\perp \x  l }{l^-} \nn \\
\apa \left.   \g^- \g^+ (q^- - l^-) \g^- \frac{2l^+}{l^-}  \right\}   \nn \\
\ata \left. \frac{\g^+ q^-  }{2p^+q^-( x_0 - x_B - {x_D}_0  + i \e)}  \right]  \nn \\
\ata 2\pi \frac{\kd [x_0 - x_B - x_L ]}{2 p^+ q^- (1- y)}. \label{O_00_2}
\eea

\nt
The $\delta$-function may be used to carry out the $p_0^+ = x_0 p^+$ integral; 
the absence of  $p_0^- $ and $p_{0_\perp}$ from the integrands 
allows for these integrals to be carried out and constrain the locations 
$y_0^+, y_{0_\perp}$ to the origin. 
Further simplifications may be carried out by noting that $\g^\pm$ anti-commutes with $\g_\perp$, while  
$\{ \g^+, \g^- \} = 2 \mbox{\bf 1}$ (\tie, twice the unit matrix in spinor space) and $\{ \g^\pm , \g^\pm \} =0 $. 
Replacing $l^- = q^- y$, in Eq.~\eqref{O_00_2}, one obtains,

\bea 
\Op^{00} &=& \kd(y_0^+) \kd^2(y_{0_\perp}) \int \frac{d y d^2 l_\perp}{(2 \pi )^3 2 y } 
e^{-i (x_B + x_L)p^+ y_0^-} p^+ \nn  \\ 
\ata \frac{g^2 4 (q^-)^2 }{(2p^+q^-)^2 4 p^+ q^- (1-y) x_L^2 }  \nn \\ 
\ata \left[  2 ( x_L p^+  - l^+ )  
 + \frac{2 l_\perp^2}{l^-} + 2 (1-y) \frac{2 l^+}{y} \right]  \nn \\
\aqa \!\!\! \kd(y_0^+) \kd^2(y_{0_\perp})  \frac{\A_s}{2\pi}  
\!\!\int \!\! \frac{dy dl_\perp^2}{l_\perp^2}  \frac{2 - 2y + y^2}{y}  . \label{O_00_3}
\eea

Reintroduction of the final state projection $\Op^{00}$  
in Eq.~\eqref{w_mu_nu_twist=2}, produces
the well known and physically clear formula 
for the differential semi-inclusive hadronic tensor with single photon 
emission in the final state,

\bea
\frac{{ dW_0^A}{\mu \nu} }{dy d l^2_\perp} &=&C_p^A 2 \pi  \sum_q Q_q^2 f^A_q (x_B+x_L) 
%
(-g_{\perp}^{\mu \nu}) \nn \\ 
\ata  \frac{\A_{s}}{2\pi}   \frac{1}{l_\perp^2}  P_{q\ra q \g }(y). \label{W_mu_nu_2}
\eea

\nt
In the equation above, $f^A_q(x_B + x_L)$ represent the parton distribution function of 
a quark with flavor $q$ and electric charge $Q_q$ in units of the electron charge 
$e$, in a nucleus with momentum $Ap$ where the parton carries a fraction $(x_B + x_L)$ of 
the momentum p,\tie,

\bea
f^A_q(x_B+x_L)  &=& A \int \frac{d y_0^-}{2\pi} e^{-i(x_B + x_L)p^+ y_0^-} \nn \\
\ata \hf \lc p| \psibar(y_0^-) \g^+ \psi(0) | p \rc .
\eea

\nt
In Eq.~\eqref{W_mu_nu_2}, the factor $P_{q\ra q \g} (y) = (2 - 2y + y^2)/y$ 
is the quark-to-photon splitting function; 
it represents the probability that a quark will radiate a photon which will carry away a fraction 
$y$ of its forward momentum. The projection $g_\perp^{\mu \nu} =g^{\mu \nu}  -  g^{\mu -} g^{\nu +} - g^{\mu +} g^{\nu -}$.

In DIS experiments, one has an experimental handle on the Bjorken variable $x_B$. In this 
article we focus on the region where $x_B \lesssim 1$. As a result, given that $l_\perp \sim \ld Q$, we may 
approximate $f(x_B + x _L) \simeq f(x_B)$.
As the parton produced immediately after the hard scattering has a vanishingly small transverse momentum, 
the transverse momentum of the final produced quark is simply the 
negative of the radiated photon's transverse momentum, 
\tie, $\vec{l}_{q_\perp} = - \vec{l}_\perp$.  As a result, the differential hadronic tensor for the 
transverse momentum distribution of the final quark is given as

\bea
\frac{d  W_0^{\mu \nu}}{dy dl_\perp^2 d^2{l_q}_\perp } &\simeq& C_p^A 2 \pi  \sum_q Q_q^2 f^A_q (x_B) 
%
(-g_{\perp}^{\mu \nu}) \nn \\ 
\ata  \frac{\A_{s}}{2\pi}   \frac{1}{l_\perp^2}  P_{q\ra q \g }(y) \kd^2 (\vec{l}_\perp + \vec{l}_{q_\perp} ). \label{d_W_0}
\eea


 \section{Gluon radiation from multiple scattering }


In this section, we start by taking the part factorized from the hard 
cross section denoted as $\mathcal{O}$ and expand it as 
a power series in the hard scale $Q^{2}$. The leading term 
was denoted as $\mathcal{O}^{00}$ in the preceding section and 
only included the process of vacuum radiation without any scattering. 
Here we consider the effect of multiple scattering on both the 
quark and the radiated gluon. As in prior attempts in this direction, 
we will not discuss the issue of factorization~\cite{Qiu:1990xx,Qiu:1990xy,col89,lqs}.
The focus will be on deriving the physical leading effect of multiple 
scattering on the process of single gluon radiation.

We start with the part denoted as $\Op_{q,p;n,m}^{N,N}$ i.e., 
the term with N scatterings on either side and with the gluon 
radiated from location $q,p$ on the quark amplitude and complex
conjugate and scattering a total of $n \leq N-q$ and $m \leq N-p$
times. This is represented in Fig.~\ref{fig2}. The Feynman integral 
for this diagram can be written down in position space, with the quark 
propagator in the amplitude (left hand side of the cut) from $y'_{i}$ to 
$y'_{i+1}$ written as, 
\bea
\mbx \!\!\!\!\!\!\!\!\!S(y_{i+1} - y_{i}) &=& 
\int \frac{d^{4} {q'}_{i+1}}{ ( 2 \pi )^{4} } \frac{ i \f {q'}_{i+1}  e^{-i {q'}_{i+1} \x  ( y'_{i+1}  - y'_{i} )  } }
{ {q'}_{i+1}^{2} + i\e } .
\eea
The quark propagators in the complex conjugate (right hand side of the cut) are identical with 
the signs of the $i \e$ and the $i$ in the numerator reversed. We then shift the momenta as 
$q'_{i+1} = q + \sum_{j=0}^{i} p'_{i} $ where $p'_{0}$ is the momentum of the original incoming 
quark which is struck with the virtual photon and the other $p'_{i}$'s are the momenta of the 
gluons off which the final out going quark scatters. Thus we replace the variable being integrated 
as $dq'_{i+1} \ra d p'_{i} $. Similar replacements are made in the complex conjugate. 
In the the gluon propagators, the replacements are made starting 
from the cut end, e.g., the $i^{\rm th}$ propagator momentum in the complex conjugate 
$Q_{i} = l - \sum_{j=i}^{m} k_{i}$ where 
$l$ is the momentum of the gluon propagator that is cut and the $k_{i}$'s are the momenta 
being brought in by the ``soft'' gluons off which the hard gluon scatters.  this 
replaces $dQ_{i} \ra d k_{i}$ and similarly in the amplitude.

Having made all these shifts in momentum we can write down the full expression for the 
Feynman integral corresponding to the diagram of Fig.~\ref{fig2}. 
We quote the full expression first and then will describe the 
various terms within.  The full expression for the final state outgoing quark and 
gluon which scatter a total of $N$ times in the amplitude and complex conjugate, 
with the gluon being produced at $q$ in the amplitude scattering $n$ times and at 
$p$ in the complex conjugate scattering $m$ times, is given as, 
\begin{widetext}
\bea
\Op^{NN}_{q,p;n,m} &=& \prod_{i=1}^{N-m} \prod_{k=1}^{N-n} d^4 y_i d^4 y^\p_k 
\frac{d^4 p_{i-1}}{(2\pi)^4} \frac{d^4 p^\p_{k-1}}{(2\pi)^4} \frac{d^4 l}{(2\pi)^4}  \frac{d^4 l_q}{(2\pi)^4} 
\prod_{i=1}^{m} \prod_{j=1}^n d^4 \zeta_i d^4 \zeta^\p_j \frac{d^4 k_i}{(2\pi)^4} \frac{d^4 k_j^\p}{( 2 \pi )^4}
\label{O_NN_raw} \\
\ata 
\tr \left[ \frac{1}{2} 
\prod_{i=0}^{p} \left\{  \g^- \frac{ - i \left\{ (\sum_{j=0}^{i} \f p_j) + \f q  \right\} }
{ [   ( \sum_{j=0}^{i} p_j  )  +  q  ]^2 - i\e}  \right\} \right. 
%
\g_{\A_0} 
\prod_{i=p}^{N-m-1} 
\left\{  \frac{   - i \left\{ ( \sum_{j=0}^{i} \f p_j )   + \f q - \f \,l + \sum_{j=1}^m \f k_j \right\} }
{ [ ( \sum_{j=0}^{i} p_j   ) + q - l + ( \sum_{j=1}^m k_j  ) ]^2 - i\e} 
\g^- \right\} \nn \\
\ata 2\pi \kd(l^2) 2\pi \kd(l_q^2) G^{\A_0 \B_0^\p}_{a_0 b_0^\p}  (m,n) \f\,l_q \nn \\
\ata \prod_{k=N-n-1}^{q} 
\left\{ \g^- \frac{ i \left\{ ( \sum_{l=0}^{k} \f p^\p_l  )  + \f q - \f \,l + \sum_{l=1}^n \f k^\p_l \right\}} 
{ [ ( \sum_{l=0}^{k} p^\p_l  )   + q -  l  + ( \sum_{l=1}^n k^\p_l ) ]^2 + i\e} \right\} 
%
\g_{\B_0}  
\prod_{k=q}^{1} \left\{  \frac{  i \left\{ (  \sum_{l=0}^{k} \f p^\p_l  )  + \f q  \right\} }
{ [   (  \sum_{l=0}^{k} p^\p_l   )  + q ]^2 + i\e} \g^- \right\} 
%
\left. \frac{ \f p^\p_0 + \f q }{ (  p^\p_0 + q)^2 + i\e} \right]  \nn \\
\ata \prod_{i=0}^{N-m-1} e^{-i p_i \x y_i } \prod_{ j=1 }^{m} e^{ -i k_j \x \zeta_j } 
\prod_{i=0}^{N-n-1} e^{i p^\p_i \x y^\p_i } \prod_{ j=1 }^{n} e^{ i k^\p_j \x \zeta^\p_j } \nn \\
%
%
\ata e^{-i y_{(N-m)} \x \left\{ l_q - \left( \sum\limits_{i=0}^{N-m-1} p_i \right)  -  q +  l  - \sum\limits_{i=1}^{m} k_i \right\} } 
e^{ i y^\p_{(N-n)} \x \left\{ l_q - \left( \sum\limits_{j=0}^{N-n-1} p^\p_j \right)  - q + l  - \sum\limits_{j=1}^{n} k^\p_j\right\} }  \nn \\
%
\ata \left\lc A \left| \prod_{i=1}^{N-m} -igt^{a_i} A^+_{a_i} (y_i) \prod_{k=N-n}^1 igt^{b_k} A^+_{b_k} (y^\p_k)  
\prod_{j=1}^m g f^{b_j a_j c_j }  A^+_{c_j} (\zeta_j) \prod_{l=1}^n g f^{c^\p_l a^\p_l b^\p_l } A^+_{c^\p_l} (\zeta^\p_l)  \
\right| A \right\rc.  \nn
\eea
The gluon line $G^{\A_0 \B_0^\p}_{a_0 b_0^\p} $ (this refers to combination of gluon propagators and scattering vertices extending 
from the amplitude side to the complex conjugate) is given as, 
\bea
\mbx\!\!\!\!\!\!\!G^{\A_0 \B_0^\p}_{a_0 b_0^\p}  (m,n)\!\!\! &=&\!\!\! \prod _{i=1}^{m} \left[ 
\frac{-i \bar{G}^{\A_{i-1} \B_{i}}_{a_{i-1} b_i} }{ \left[ l - \sum_{j=i}^m k_j \right]^2 - i\e  } 
(-g_\perp)_{\B_i \A_i} 2l^- 
%
\right] 
(-g_\perp)_{\A_m \B^\p_n} \kd^{a^m {b^\p}^n} 
 \prod_{k=n}^1 \left[  2l^- 
(-g_\perp)_{\B^\p_i \A^\p_i}  
\frac{i \bar{G}^{\A^\p_{i} \B^\p_{i-1} }_{a_i b_{i-1}}}
{ \left[ l - \sum_{l=k}^n k_l^\p \right]^2 + i \e } \right]\!. 
\label{glue}
\eea
\begin{figure}[htbp]
\resizebox{3in}{3in}{\includegraphics[2in,0in][8in,6in]{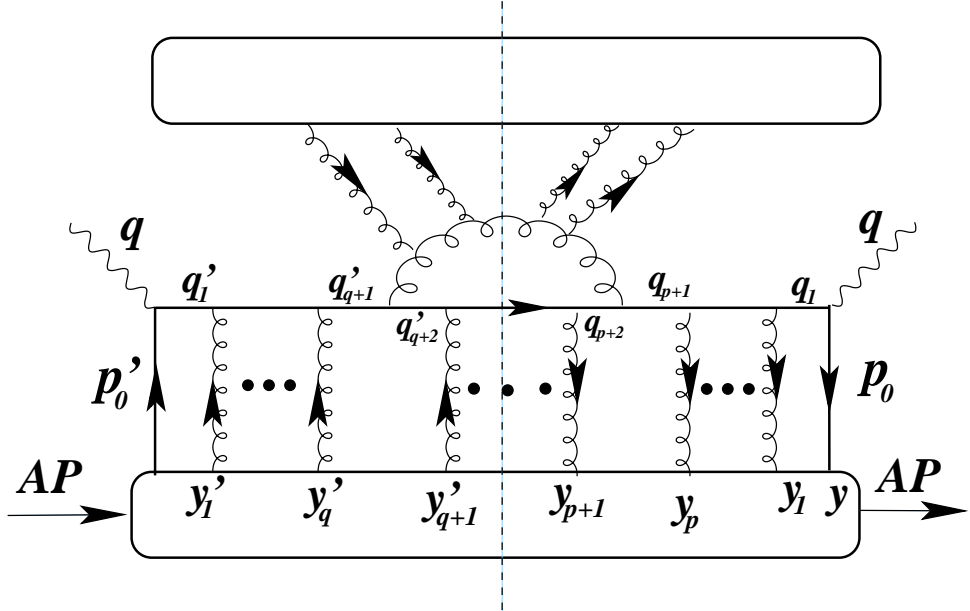}} 
    \caption{ DIS on a nucleus with a produced quark and an emitted gluon scattering $N$ times.}
    \label{fig2}
\end{figure}

\end{widetext}
The Feynman diagram corresponding to this equation is Fig.~\ref{fig2}. 
The first line of Eq.~\eqref{O_NN_raw} contains a series of integrals over coordinates and momenta. 
The coordinates $y_i$, where $1< i < N-m$, represent the locations where the 
quark scatters in the amplitude; the coordinates $y^\p_k$, where $1< k < N-n$, represent the scattering 
of the quark in the complex conjugate. The momenta $p_{i-1}$, are the incoming momenta into the 
quark line with $p_0$ being the momentum of the incoming quark which is struck by the virtual photon. 
The remaining momenta from $p_1$ to $p_{N-m-1}$ are momenta of the gluons off which the quark 
scatters. The same is true for the $p^\p_k$ momenta in the complex conjugate. The coordinates 
$\zeta_i$ ($\zeta^\p_j$), where $1<i<m$ ($1<j<n$) represent the locations where the emitted gluon 
scatters in the amplitude
(complex conjugate). The momenta carried by each of these insertions is characterized by $k_i$ ($k^\p_j$)
where $1<i<m$ ($1<j<n$) in the amplitude (complex conjugate). The two remaining momenta $l$ and $l_q$ 
represent the final on-shell momenta of the exiting quark and gluon as in the previous section. 

The second line of Eq.~\eqref{O_NN_raw} represents the quark propagators on the 
amplitude side. The third line contains the cut propagators and the full set of 
gluon propagators from $\A_0$ to $\B_0^\p$ in one expression  $G^{\A_0 \B_0^\p}_{a_0 b_0^\p} $ which 
is decomposed in Eq.~\eqref{glue}. The Greek superscripts represent the vector indices where the gluon is 
first emitted from the quark line and the Latin subscripts represent the color at this emission. 
The fourth line contains the quark propagators on the 
complex conjugate side of the cut. The fifth line contains the phase factors for all but the cut quark propagator. 
This last phase factor is separately expressed in the sixth line. The seventh line contains the expectation in the 
nucleus for the $N+N$ gluon field operators in the amplitude and complex conjugate. Here $| A  \rc$ represents 
the nuclear state~\cite{f2}.

Eq.~\eqref{glue}, contains all the gluon propagators for the emitted gluon which 
undergoes $m$ scatterings in the amplitude and $n$ scatterings in the complex 
conjugate,
where, $\bar{G}^{\A_{i-1} \B_{i}}_{a_{i-1} b_i} $ represents the numerators 
of  the gluon propagators 
after the emission both on the amplitude and complex conjugate side. In the rest of 
this section we will discuss various simplifications that arise due to our choice of 
kinematics for the radiated gluon: $l_\perp \sim \ld Q$ and $l^- = yq^- \sim Q$ where 
$Q$ is the hard scale in the DIS process.

Eqs.~(\ref{O_NN_raw},\ref{glue}) are not completely general and have a set 
of approximations already included.
While calculations are carried out in the negative light-cone gauge $A^-_{a_i} = 0$, 
there remain three other components of the gluon field 
in covariant products such as  $\g_{\mu_i} A^{\mu_i}_{a_i}$ which arise with each 
quark-gluon vertex. 
As shown in a recent study~\cite{Idilbi:2008vm}, as well as in Appendix A, in the case of the Breit frame, 
the dominant component of the vector potential (in $A^- = 0$ gauge), for gluons radiated from partons 
with collinear momenta in the $(+)$-direction, is the ($+$)-component which scales as
$A^+ \sim \ld^2 Q$ compared to the ($\perp$)-component which scales as 
$A_\perp \sim \ld^3 Q$.

The numerator of the terms in the second and fourth line of Eq.~\eqref{O_NN_raw} have 
the general form $\ldots \g_{\mu_i} q_i^{\mu_i} \g_{\nu_i} A^{\nu_i}_{a_i} t_{a_i} \ldots$. 
While this includes all four components of the quark momentum and the gauge field, the 
$(-)$-component of the quark momenta is $q_i^- = q^- + \sum_{j} p_j^- \simeq q^-\sim Q$ 
and dominates over the ($\perp$) and 
$(+)$-components which scale as $\ld Q$ and $\ld^2 Q$ respectively. 
This is also true of the gauge field as mentioned above. As a result, 
the leading term in each product of quark propagator and gluon insertion is 
$ \ldots \g^+ q^- \g^- A^+_{a_i} t^{a_i} \ldots$ if the quark propagator 
is situated before the radiation of the outgoing gluon and is 
$\ldots \g^+ (q^- - l^-)  \g^- A^+_{a_i} t^{a_i} \ldots$ 
for quark propagators which appear after the radiated gluon. 
This is identical to the case of photon radiation 
from a multiply scattering quark produced in DIS~\cite{Majumder:2007ne}. 

The major difference in this article from Ref.~\cite{Majumder:2007ne}, besides the scaling 
relations used, is the scattering encountered by the radiated gluon. This is somewhat non-trivial 
in $A^- = 0$ gauge. We begin by elucidating the various propagator and three gluon vertex structures. 
We will only consider three gluon vertices as four gluon vertices indicate a scattering of two gluon 
fields at the same location. As pointed out in Ref.~\cite{Majumder:2007hx}, such contributions 
are not length enhanced and are thus suppressed. From the effective theory perspective, 
such contributions are not contained in the leading Lagrangian~\cite{Idilbi:2008vm}.

The gluon propagator in $A^-=0$ gauge for a gluon with 
four-momentum $l$ contains the constant unit vector $n^{\mu} = (1,0,0,0)$ and is given as (suppressing
color indices) 
\bea
G^{\mu \nu} (l)&=& \frac{i \bar{G}^{\mu \nu}}{l^2 + i \e} 
= i \frac{ -g^{\mu \nu}  + \frac{l^\mu n^{\nu} + l^\nu n^\nu}{l\x n} }{l^2 + i \e}
\eea
All components where either of the Lorentz indices is ($-$), are vanishing, i.e., 
\bea
G^{- +} &=& G^{+ -} = i \frac{ -1 + 1 }{l^2} = 0, \nn \\
G^{- -} &=& G^{\perp -} = 0.
\eea
The leading components (where the numerator has the largest power of $\lambda$) 
are those where both Lorentz indices are transverse, i.e.,
\bea
(G_\perp)_{\mu \nu} \equiv G_{\perp_\mu \perp_\nu} = i \frac{ - g_{\perp_\mu \perp_\nu} }{ l^2 + i\e} 
= i \frac{ - {g_\perp}_{\mu \nu} }{ l^2 + i\e} .
\eea
The meaning of the notation $l_{\perp_\mu} \equiv (l_\perp)_\mu$ is one where the index, which 
may be referred to as $\mu$ or $\perp_\mu$ can 
only assume values $\mu = \perp_\mu = 1,2$, 
and $(l_\perp)^\mu = l^{\perp_\mu} = - l_{\perp_\mu} = -(l_\perp)_\mu$ 
and $ (g_\perp)_{\mu \nu} \equiv g_{\perp_\mu \perp_\nu} = g^{\perp_\mu \perp_\nu} \equiv (g_\perp)^{\mu \nu} = - \kd^{ \perp_i \perp_j}$.
The $\perp \perp$ components are followed by the ($+ \perp$) and the ($++$) 
components, where
\bea
G^{+ \perp_i} &=& i\frac{\bar{G}^{+ \perp_i}}{l^2 + i\e} = i \frac{l^{\perp_i}}{(l^-) [ l^2 + i \e] },   \nn \\
G^{++} &=& i \frac{ \bar{G}^{+ +}}{l^2 + i\e} =  i \frac{ 2 l^+  }{l^- [l^2 + i\e] } .
\eea

The three gluon vertex contracted with the gauge field may also be simplified as, 
\bea
\mbx\!\!\!\!\! && \Gamma_{\B \A \g}^{bac} (l,k) A_c^\g(k) = gf^{bac} 
\label{3_gluon_vert}\\
\mbx\!\!\!\!\!  
\ata [g_{\B \A} (k - 2l)\x A_c + {A_c}_\A (l -  2k)_\B  + {A_c}_\B (l + k)_\A ] \nn \\
\mbx\!\!\!\!\!  
&\simeq&gf^{bac} g_{\B \A} (- 2l^-) A_c^+ +  {A_c}_\A l^- g^+_\B + {A_c}_\B l^- g^+_\A. \nn 
\eea
When the three gluon vertex is used in combination with gluon propagators on 
either side, the dominant term in terms of counting powers of $\ld$ 
from the contraction of Lorentz indices is 
the first term on the third line of Eq.~\eqref{3_gluon_vert}, i.e.,
\bea
& & \ldots \bar{G}^{\A_{i-1} \B_i}_{a_{i-1} b_i} 
\Gamma_{\B_i \A_i \g_i}^{b_i a_i c_i} A_{c_i}^{\g_i} 
\bar{G}^{\A_{i} \B_{i+1}}_{a_{i} b_{i+1} } \ldots \label{3_gluon_with_prop} \\
&=& \ldots  \left(- {g_\perp}^{ \A_{i-1}  \B_i } \right)
g f^{a_{i-1} b_{i+1} c_i} A^+_{c_i} 2l^- \nn \\
\ata \left( -{g_\perp}_{\B_i \A_i} \right) 
\left( -{g_\perp}^{ \A_{i}  \B_{i+1} } \right) \ldots .\nn
\eea

The approximation of replacing the numerators of all the gluon 
propagators with the transverse projector ${g_\perp}_{\mu \nu}$
is consistent for all except the first and the last propagators which
connect with the quark line. Here the Lorentz index that contracts 
with the $\g$-matrix may assume both a $+$ or a $\perp$ 
component yielding the same power counting for the complete
expression. As a result, in Eq.~\eqref{glue}, the numerators of the 
first and last gluon propagators may contain either a
 $(\bar{G}^{+ \perp} )^{\mu \nu}$ or a ${g_\perp}^{\mu \nu}$. The numerators 
of all other gluon propagators contain solely the sum over transverse 
polarizations to obtain the leading power counting.

As a result of the above simplifications, the $A^{\mu}$ fields in Eq.~\eqref{O_NN_raw} 
have all been replaced by $A^+$. The contracting $\g^\mu$ matrices  are 
replaced by simply $\g^-$. The momentum dependent part of the three gluon vertices 
have been replaced by simply $2l^-$ as indicated by Eq.~\eqref{3_gluon_with_prop}.
There remain the simplifications on the numerators of the quark propagator, which
essentially amount to replacing all the numerators with $\g^+q^-$ expect for 
the propagators that surround the gluon emission vertex. The contraction with either 
a $ ( \bar{G}^{+ \perp} )^{\mu \nu}$ or a $g_\perp^{\mu \nu}$ leads to three 
terms from the numerator of the three propagators connecting to the gluon emission 
vertex, 
\bea
V^{\nu} &\simeq& \g^+ q^- \g_\mu  \frac{ n^\mu \lp l_\perp - \sum_{j=1}^m k_\perp^j \rp^\nu }{l^-} \g^+ (q^- - l^-)  \\
&+& \g_\perp^\A \lp \sum_{i=0}^p p^i_\perp \rp_\A {\g_\perp}_\mu (-g_\perp^{\mu \nu})  \g^+ (q^- - l^-) \nn \\
&+& \g^+ q^- {\g_\perp}_\mu (-g_\perp^{\mu \nu}) \g_\perp^\A \lp \sum_{i=0}^p p_\perp^i + 
\sum_{j=1}^m k^j_\perp - l_\perp \rp_\A  . \nn
\eea
A similar structure arises on the complex conjugate side as well. Invoking all these simplifications 
leads to a considerable simplification of the numerator structure of Eq.~\eqref{O_NN_raw}. 

Following the stated rules of power counting, the denominator of the various propagators 
in Eqs.~(\ref{O_NN_raw},\ref{glue}) may be simplified and expressed in terms of the usual 
momentum fractions~\cite{HT,Majumder:2007ne}. For the first set of quark denominators, 
up to the gluon emission vertex, we introduce the fractions, 
\bea
x_D^i = \frac{ | p^i_\perp |^2 + 2 \sum_{j=0}^{i-1} p^i_\perp \x p^j_\perp }{2 p^+ q^- } \sim \ld^2. 
\label{x_D_i}
\eea
While written without vector attributes, it should be understood that $p_\perp^i$ for all $i$ 
are two dimensional vectors transverse to the direction of propagation of the jet. 
Using this, we may write the $i^{\rm th}$ denominator as, 
\bea
D_{i \leq p} = 2p^+q^- \left[ \left( \sum_{j=0}^i x_i  - x^i_D \right) - x_B  \right],
\eea
where $x_i = p_i^+/p^+  \sim \ld^2$ and is thus included in the denominator. Ignored are 
terms which depend on $p_i^-$ which introduce $\ld^2 Q$ corrections to $q^- \sim Q$.
For the denominators beyond the emitted photon, besides the fractions $x_D^i$ which are 
defined similar to Eq.~\ref{x_D_i}, we introduce the momentum fractions which connect the 
final outgoing transverse momentum of the radiated gluon with the incoming 
transverse momentum which strikes the hard quark,
\bea
x_L^i = \frac{ | l_\perp|^2   -  2 y l_\perp \x \sum_{k=0}^i p^k_\perp }{ 2 p^+ q^- y (1 - y) }.
\eea
Another set of transverse momentum fractions arise from the combination of the incoming transverse 
momenta striking the gluon line and those striking the quark line,
\bea
( x^m_T )^i = \frac{ 2 p_\perp^i  \x \sum_{j = 1}^m k_\perp^j  }{ 2 p^+ q^- }.
\eea
Finally there are the momentum fractions that connect the transverse momentum of the 
final radiated gluon with the soft transverse momenta striking it, 
\bea
(z_L^m)_i = \frac{ 2 l_\perp \x  k_\perp^i  -  | k_\perp^i |^2 
- k_\perp^i \x  \sum_{j=i+1}^m k^j_\perp    }{ 2 p^+ q^-}.
\eea
The forward momentum fractions of these incoming gluons denoted as, 
\bea
z_i = \frac{ k_i^+ }{p^+}.
\eea
With these fractions, the denominator of the $i^{\rm th }$ quark propagator 
after the gluon emission can be expressed as, 
\bea
D_{i>p} &=& 2 p^+ q^- ( 1 - y ) \lb  \lp \sum_{j=0}^i  x_j - \frac{x^j_D + (x^m_T)^j }{1-y}   \rp \right.  \nn \\
&-&  \left. x_B - x_L^i + \lp \sum_{k=1}^m  z_k + \frac{ (z_L^m)_k }{ 1 - y }   \rp \rb .
\eea
The denominator of the $i^{\rm th}$ gluon propagator has a much simpler expression in terms of 
these momentum fractions, 
\bea
D^g_i &=& -2 p^+ q^- y \lb \sum_{j=i}^m z_j - \frac{(z_L^m)_j }{y}   \rb .
\eea

To Substitute the above simplifications, requires that we separate the trace over spinor indices 
contracted with the vector indices in the numerators of the gluon propagators in a numerator factor $\mathcal{N}$, and 
the remaining denominators of all propagators in a denominator factor $\mathcal{D}$, which are coupled by the set of 
integrals over position and momentum, i.e.,
\bea
\Op^{NN}_{q,p;n,m} = \int \md (y)  \md(p)  \mathcal{N} \x \mathcal{D},
\eea
where, $\md(y)$ represents all position integrals on the amplitude and complex
conjugate side of the matrix element and $\md(p)$ represents all the integrals over momenta. 

The numerator factor, containing all the $\g$ matrices from the quark line and all metric tensors from the 
gluon line, keeping only the leading terms in $\ld$, simplifies as,
\begin{widetext}
\bea
\mathcal{N} &=& - \tr \left[ \frac{1}{2} (\g^- \g^+ q^-)^{p} \g^-  \left\{
\g^+ q^- \g_\mu  \frac{ n^\mu \left( l_\perp - \sum_{j=1}^m k_\perp^j \right)^\nu }{l^-} \g^+ (q^- - l^-)  \right. \right.
+ \g_\perp^\A \left( \sum_{i=0}^p p^i_\perp \right)_\A {\g_\perp}_\mu (-g_\perp^{\mu \nu})  \g^+ (q^- - l^-) \nn \\
&+& \g^+ q^- {\g_\perp}_\mu (-g_\perp^{\mu \nu}) \g_\perp^\A \lp \sum_{i=0}^p p_\perp^i + 
\left. \sum_{j=1}^m k^j_\perp - l_\perp \rp_\A   \right\} \{ \g^- \g^+ (q^- - l^-) \}^{N-m-p-1}  \g^-\g^+ (q^- - l^-) \nn \\
\ata \{ \g^-\g^+ (q^- - l^-) \}^{N-n-q-1} \g^-\left\{
\g^+ ( q^- - l^-) \g_\ro  \frac{ n^\ro \lp l_\perp - \sum_{j=1}^n {k^\p}_\perp^j \rp_\nu }{l^-} \g^+ q^-   \right.  \\
&+& \g^+ (q^- - l^-) \g_\perp^\ro (-g_\perp)_{\ro \nu}  \g_\perp^\B \lp \sum_{i=0}^q {p^\p}^i_\perp \rp_\B 
+ \g_\perp^\B \lp \sum_{i=0}^q {p^\p}_\perp^i \left. \left.
+ \sum_{j=1}^n {k^\p}^j_\perp - l_\perp \rp_\B \g_\perp^\ro (-g_\perp)_{\ro \nu}    
\g^+ q^-  \right\} (\g^- \g^+ q^- )^{q} \rb \nn \\
&=& - \frac{1}{2} \tr \left[  (2q^-)^p \g^-  \left\{
\g^+ q^- \g_\mu  \frac{ n^\mu \left( l_\perp - \sum_{j=1}^m k_\perp^j \right)^\nu }{l^-} \g^+ (q^- - l^-)  \right. \right.
+ \g_\perp^\A \left( \sum_{i=0}^p p^i_\perp \right)_\A {\g_\perp}_\mu (-g_\perp^{\mu \nu})  \g^+ (q^- - l^-) \nn \\
&+& \g^+ q^- {\g_\perp}_\mu (-g_\perp^{\mu \nu}) \g_\perp^\A \left( \sum_{i=0}^p p_\perp^i + 
\left. \sum_{j=1}^m k^j_\perp - l_\perp \right)_\A   \right\}  \{ 2 (q^- - l^-) \}^{N-m-p-1}  2(q^- - l^-) \nn \\
\ata \{ 2 (q^- - l^-) \}^{N-n-q-1} \g^-\left\{
\g^+ ( q^- - l^-) \g_\ro  
\frac{ n^\ro \left( l_\perp - \sum_{j=1}^n {k^\p}_\perp^j \right)_\nu }{l^-} \g^+ q^-   \right.  \\
&+& \g^+ (q^- - l^-) \g_\perp^\ro (-g_\perp)_{\ro \nu}  \g_\perp^\B \left( \sum_{i=0}^q {p^\p}^i_\perp \right)_\B 
+ \g_\perp^\B \left( \sum_{i=0}^q {p^\p}_\perp^i \left. \left.
+ \sum_{j=1}^n {k^\p}^j_\perp - l_\perp \right)_\B \g_\perp^\ro (-g_\perp)_{\ro \nu}    
\g^+ q^-  \right\} (2q^-)^{q} \rb \nn 
\eea
%
The entire set of transverse components of the metric tensor ($-{g_\perp}^{\mu \nu}$) that arise from the 
propagators and vertices of the radiated gluon line lead to the contraction between the $\nu$ component of the 
gluon emission vertex in the amplitude with the same component in the complex conjugate. The overall minus sign 
is due to the fact that there is one more propagator in the gluon line than vertex i.e., the cut propagator.

The equation above for $\mathcal{N}$ does not contain any ($+$) or ($-$) components of the incoming momenta: $p_i$, $p^\p_i$, $k_i$, $k^\p_i$. The only non-transverse components are the large ($-$) components of the 
outgoing quark ($q^-$) and radiated gluon ($l^-$). This allows for the integrations over all ($+$) and ($-$) 
components of the incoming momenta to be carried out solely by considering the denominators of the 
various propagators. These set of integrations are carried out in the next section. 

The remaining expression for the numerator of $\Op^{NN}_{q,p;n,m}$ can be further simplified to obtain a 
more suggestive form. Isolating the coefficient $C_{L}$ of 
$(\vl_\perp - \sum_{i=1}^m \vk^i_\perp )\x (\vl_\perp - \sum_{j=1}^n \vec{k^\p}^j_\perp  ) $ 
yields four terms, which may be summed to obtain, 
\bea
C_L &=& \frac{ (2 q^-)^{2N - m -n + 1}}{y} ( 1 - y )^{ 2 N -m -n - p -q - 1 } 
\lt[ \frac{4}{y} (1-y)^2 + 2 (1-y) + 2 (1-y) + 2y \rt]  \\
&=& \frac{ (2 q^-)^{2N - m -n + 1}}{y} ( 1 - y )^{ 2 N -m -n - p -q - 1 }  \lt[ \frac{2}{y} \lt( 2 - 2y + y^2 \rt)  \rt]
= \frac{ (2 q^-)^{2N - m -n + 1}}{y} ( 1 - y )^{ 2 N -m -n - p -q - 1 } 2 P(y). \nn  \label{coeff_1}
\eea
The coefficients of $ y \sum_{i=1}^{p} \vp^{\,i}_\perp  \x \left(  \vl_\perp - \sum_{j=1}^n \vec{k^\p}^j_\perp  \right) $, 
$ \lt( \vl_\perp - \sum_{i=1}^m \vk^i_\perp \rt) \x y \sum_{j=1}^q { \vec{p^\p} }^j_\perp $ and 
$y^2 \sum_{i=1}^{p} \vp^{\,i}_\perp  \x  \sum_{j=1}^q { \vec{p^\p} }^j_\perp$ are similar
and may be included with the contributions containing $C_L$ to obtain the general numerator as 
\bea
\mathcal{N} &=& \frac{ (2 q^-)^{2N - m -n + 1}}{y} ( 1 - y )^{ 2 N -m -n - p -q - 1 } 2 P(y) 
\lt( \vl_\perp - \sum_{i=1}^m \vk^i_\perp - y \sum_{i=1}^{p} \vp^{\,i}_\perp  \rt) 
\x 
\lt( \vl_\perp - \sum_{j=1}^n \vec{k^\p}^j_\perp  - y \sum_{j=1}^q { \vec{p^\p} }^j_\perp \rt) \label{numerator}.
\eea
Ignoring the factors of $y, (1-y)$ and $q^-$ on the right hand side, the remaining expression for the 
numerator has a very simple 
interpretation. While $P(y)$ is simply the regular vacuum splitting function, the two terms, 
$\lt( \vl_\perp - \sum_{i=1}^m \vk^i_\perp - y \sum_{i=1}^{p} \vp^{\,i}_\perp  \rt) $ and 
$\lt( \vl_\perp - \sum_{j=1}^n \vec{k^\p}^j_\perp  - y \sum_{j=1}^q { \vec{p^\p} }^j_\perp \rt)$ 
are the transverse momentum of the radiated gluon immediately after the radiative vertex in 
the amplitude and complex conjugate respectively. In the next section we will simplify the 
denominator by carrying out a series of integrations over the light-cone components of the 
incoming gluon momenta. The numerator factor in Eq.~\eqref{numerator} does not contain these
components and thus plays no role in these integrations.


\section{The Poles structure for single gluon emission from N scattering}


In this section, the integrals over the light cone components of the incoming 
gluon momenta will be carried out. As mentioned in the preceding section, for this part of the integration, 
we only need to consider the denominators of all the propagators. The phase 
factors for the last scatterings on the quark at locations $y_{(N-m)}$ may be 
simplified, by the 
introduction of an $(N-m)^{\rm th}$ momentum, using the four $\kd$-function as

\bea
1 &=& \int d^4 p_{(N-m)} 
\kd^4 \left( l_q - \left( \sum\limits_{i=0}^{N-m-1} p_i \right)  -  q +  l 
- \sum\limits_{i=1}^{m} k_i  - p_{(N-m)}\right). 
\eea
Using this the $\pm$ components of this four dimensional $\kd$-function and the 
various momentum fractions, the argument of the $\kd(l_q^2)$ function may 
be simplified as, 
\bea
&&\int d l_q^+ d l_q^- \kd(l_q^2) \kd^2 \left( l_q^\pm - \sum_{i=0}^{N-n} p_i^\pm - q^\pm + l^\pm - \sum_{i=1}^m k_i^\pm   \right)
=  \int d l_q^+ d l_q^- \frac{ \kd \left( l_q^+ - \frac{{l_q}_\perp^2}{ 2 l_q^- } \right) }{ 2 l_q^-} 
\kd^2 \left( l_q^\pm - \sum_{i=0}^{N-n} p_i^\pm - q^\pm + l^\pm - \sum_{i=1}^m k_i^\pm   \right) \nn \\
&\simeq& \frac{1 }{2 p^+ q^- (1 - y)} \kd \left( \sum_{i=0}^{N-m} x_i - \frac{x_D^i + (x_T^m)^i }{(1-y)} 
-   x_B - x_L^{(N-m)} + \sum_{j=1}^m z_j + \frac{(z^m_L)_j}{(1 - y)}  \right).
\eea
We also perform the integration over $l^+$ using the on-shell delta function, 
\bea
\frac{d^4 l}{(2 \pi)^4} 2\pi \kd (l^2) &=&  \int \frac{dl^- d^2 l_\perp}{ (2 \pi)^3} \frac{1}{l^-} .
\eea
We make the substitution, $l^- = q^- y$ and express all further expressions in terms of the radiated momentum 
fraction $y$.
This simplifies the last two phase factors in the expression for $\Op^{NN}_{q,p;n,m}$ at locations $y_{(N-m)}$ and 
$y^\p_{(N-n)}$. Including only the denominators, $\kd$-functions (associated factors of $2\pi$ and $i$) 
and phase factors from Eq.~\ref{O_NN_raw}, we obtain, 
%
%
\bea
\md^{NN}_{q,p;n,m} \!\!\!\!&=&\!\!\!\! \prod_{i=1}^{N-m} \prod_{k=1}^{N-n} d^4 y_i d^4 y^\p_k 
\frac{ dx_{i-1} d^3 p_{i-1} }{ (2\pi)^4 }  \frac{ dx^\p_{k-1}  d^3 p^\p_{k-1} }{ (2\pi)^4 } 
\frac{ dy d^2 l_\perp }{ (2\pi)^3 y }  \frac{ d x_{(N-m)} d^3 p_{(N-m)} }{ (2 \pi)^4 }
\prod_{i=1}^{m} \prod_{j=1}^n d^4 \zeta_i d^4 \zeta^\p_j 
\frac{d z_i d^3 k_i}{ (2\pi)^4 } \frac{ d z^\p_j d^3 k_j^\p }{ ( 2 \pi )^4 }  
\nn \\
\ata d^2 {l_q}_\perp  \kd^2 \left( {l_q}_\perp + l_\perp - \sum_{i=0}^{(N-m)} p^i_\perp - \sum_{j=1}^m k^j_\perp \right) 
\prod_{i=0}^p \left\{  \frac{ -i }
{ 2q^- \left[   \lp \sum_{j=0}^{i} x_i - x_D^i  \rp  - x_B   - i\e   \right]  }   \right\}  \nn \\
\ata \prod_{ i=p }^{ N-m-1 } 
\left\{  
\frac{ -i } { 2q^- (1-y) \left[ \left( \sum_{j=0}^{i} x_j  - \frac{x_D^j + (x_T^m)^j }{(1-y)} \right)
- x_B - x_L^i + \left( \sum_{j=1}^m z_j + \frac{(z^m_L)_j}{ 1 -  y }  \right)  - i \e \right] } 
\right\} \nn \\
\ata   	
\frac{1}{(p^+)^2}\frac{2\pi}{2 q^- (1 - y)} \kd  \left( \sum_{i=0}^{N-m} x_i - \frac{x_D^i + (x_T^m)^i }{(1-y)} 
-  x_B - x_L^{(N-m)} + \sum_{j=1}^m z_j + \frac{(z^m_L)_j}{ 1 - y }  \right) \nn \\
%
\ata \prod_{k=N-n-1}^{q} 
\left\{  
\frac{ i } { 2q^- (1 - y)\left[ \left(  \sum_{l=0}^{k}  x^\p_l - \frac{ {x_D^\p}^l - ({x^\p}^m_T)^l }{ 1- y }  \right)   
- x_B -  {x^\p}^k_L  +  \left( \sum_{l=1}^n  z^\p_l  + \frac{ {z^\p}^l_L  }{1-y} \right)   + i\e  \right] } 
\right\} \nn \\
%
\ata \prod_{k=q}^{0} \left\{  
\frac{ i }{  2q^- \left[   \left(  \sum_{l=0}^{k} x^\p_l - {x^\p}_D^l   \right)  - x_B  + i \e \right]   }  
\right\} \nn \\
\ata \prod_{i=1}^m \frac{-i}{-2q^-y \left[ \sum_{k=i}^m z_k - \frac{(z_L^m)_k}{y} + i\e  \right]} 
\prod_{j=1}^n \frac{i}{-2q^-y \left[ \sum_{l=j}^n z_l - \frac{(z_L^n)_l}{y} - i \e\right]} \nn \\
\ata \prod_{i=0}^{N-m} e^{-i p_i \x ( y_i  - y^\p_{(N-n)} ) } \prod_{ j=1 }^{m} e^{ -i k_j \x ( \zeta_j - y^\p_{(N-n)} ) } 
\prod_{i=0}^{N-n-1} e^{i p^\p_i \x ( y^\p_i - y^\p_{(N-n)} )  } \prod_{ j=1 }^{n} e^{ i k^\p_j \x 
( \zeta^\p_j - y^\p_{(N-n)} )  } .
%
\label{O_NN_denom_1} 
\eea
\end{widetext}
The extra factor of $(1/2p^+)^2$ is due to the fact that there are two extra propagators than 
integrals over ($+$)-components of momentum. The gluon propagators have an over all negative 
sign compared to the quark propagators. This is due to convention of calculating the momentum in 
the $i^{\rm th}$ gluon propagator in the amplitude as, 
\bea
&& \frac{i}{ \left[ l - \sum_{j=i}^q k^j  \right]^2  + i \e } \nn \\
&\simeq& \frac{i}{ \left[  - 2l^- \sum_{j=i}^n k_j^+ + l_\perp \x \sum_{j=i}^n k_\perp^j 
- \left| \sum_{j=i}^n k_\perp^j \right|^2  + i \e \right]} \nn \\
&=& \frac{i}{ -2p^+ q^- y\left[ \sum_{j=i}^n z_j - \frac{ (z_L^m)_j }{y}  - i \e \right] }.
\eea

The reader will note that besides the phase factors, the ($-$)-components of the 
incoming gluon momenta have been neglected from the integrand. As a result, 
we can integrate over the ($-$)-components of all the incoming gluon momenta 
isolating the entire process on the negative light cone, e.g.,
\bea
\int\frac{d p_i^-}{2\pi} e^{- i p_i^- ( y_i^+ - {y^\p}_{(N-m)}^+  ) } %
&=&\kd ( y_i^+ - {y^\p}_{(N-m)}^+ ).  
\eea
Like wise integrating over the set of momenta ${p_{i}^{\p}}^{-}$, $k_{i}^{-}$ and ${k_{i}^{\p}}^{-}$ 
yields the sets of $\kd$-functions 
[$\kd ( {y^\p}_i^+ - {y^\p}_{(N-m)}^+ ),   \kd ( \zeta_i^+ - {y^\p}_{(N-m)}^+ ), \kd ( {\zeta^\p}_i^+ - {y^\p}_{(N-m)}^+ )$  ] that constrains every vertex to the ${y^{\p}}_{N-m}^{+}$ surface. 
The $\kd$-function that arises from the integration over ${p_{0}^{\p}}^{-}$, equates, ${y^{\p}}_{N-m}^{+}$
and thus all other ($+$)-locations to the origin (${y^{\p}}^{-}_{0} = 0$). 
This removes all negative light-cone components of momentum from the derivation. It should 
be pointed out that, in the higher twist formalism, these components are responsible
for elastic energy loss~\cite{Majumder:2008zg,Qin:2007rn}. By integrating out these components 
we specialize to the case of only radiative energy loss. In reality, as is obvious from 
the formulation of the Feynman integral of Eq.~\eqref{O_NN_raw}, elastic and radiative 
loss are coupled, with each contributing to the magnitude of the other. We leave 
the coupled calculation of elastic and radiative loss to a future effort.

The integrations over the momentum fractions $x_i$, $x^\p_i$, $z_i$ and $z^\p_i$ requires 
a certain amount of care as regards the order in which these have to be performed in order 
to have the simplest pole structure. 
We focus only on the remaining light-cone components of the phase factor, 
\bea
\Gamma^+ &=& \prod_{i=0}^{N-m} e^{-i x_i p^+ ( y_i^-  - {y^\p}^-_{(N-n)} ) } 
\prod_{ j=1 }^{m} e^{ -i z_jp^+ ( \zeta^-_j - {y^\p}^-_{(N-n)} ) } \nn \\
\ata \prod_{i=0}^{N-n-1} e^{i x^\p_i  p^+ ( {y^\p}^-_i - {y^\p}^-_{(N-n)} )  } 
\prod_{ j=1 }^{n} e^{ i z^\p_j p^+ ( {\zeta^\p}^+_j - {y^\p}^+_{(N-n)} )  } . \nn \\
\eea
The remaining delta function over the ($+$)-components 
of the momentum may be used to set the last momentum fraction $x_{(N-m)} ( \equiv  p_{(N-m)}^+/p^+)$.   
Similar to the case of photon production and multiple scattering 
in Refs.~\cite{Majumder:2007ne} and~\cite{Majumder:2007hx}, 
we obtain, 
\begin{widetext}
\bea
x_{(N-m)} &=& x_B + x_L^{(N-m)} - \left( \sum_{k=1}^m z_k +  \frac{(z_L^m)_k}{ 1-y } \right) 
+ \sum_{i=0}^{N-m} \frac{x_D^i + x_T^i}{ 1- y }  - \sum_{i=0}^{N-m-1} x_i .
\eea
This converts the light-cone phase factor to, 
\bea
\Gamma^+ &=& \prod_{i=0}^{N-m} e^{-i x_i p^+ ( y_i^-  - y^-_{(N-m)} ) } 
\prod_{ j=1 }^{m} e^{ -i z_jp^+ ( \zeta^-_j - y^-_{(N-m)} ) } 
\prod_{i=0}^{N-n-1} e^{i x^\p_i  p^+ ( {y^\p}^-_i - {y^\p}^-_{(N-n)} )  } 
\prod_{ j=1 }^{n} e^{ i z^\p_j p^+ ( {\zeta^\p}^-_j - {y^\p}^-_{(N-n)} )  }  \nn \\
\ata \exp \left[ -ip^+ \left( y_{(N-m)}^- - {y^\p}^-_{(N-n)} \right)  \left\{  x_B + x_L^{(N-m)}  
+ \sum_{i=0}^{N-m} \frac{x_D^i + x_T^i}{ 1- y } - \sum_{k=1}^m \frac{(z_L^m)_k}{ 1-y } \right\} \right]. 
\eea

The remaining integrations over momentum fractions involves contour integrations around the 
various poles of the $2N+2$ propagators. The simplest means to carry this out is to start by integrating 
the penultimate momentum fractions on either side of the cut quark line, $x_{(N-m-1)}$ and ${x^\p}_{(N-n-1)}$
and proceed towards the lower momentum fractions. In this way, the fraction being integrated over appears in 
at most one propagator. The integration over $x_{(N-m-1)}$ requires the contour to be closed at $+ i \infty$ and, 
as a result, the positions $y^-_{(N-m)}$ and $y^-_{(N-m-1)}$ need to be ordered with $y^-_{(N-m)} > y^-_{(N-m-1)} $. 
The isolated residue of this integration yields, 
\bea
&& \int \frac{dx_{(N-m-1)}}{2\pi} 
\frac{ e^{-ix_{(N-m-1)} \x p^+ \lp y_{(N-m-1)}^- - y_{(N-n)}^- \rp }} 
{ \left(\sum_{j=0}^{N-m-1} x_j  - \frac{x_D^j + (x_T^m)^j }{(1-y)} \right)
- x_B - x_L^{(N-m-1)} + \left( \sum_{j=1}^m z_j + \frac{(z^m_L)_j}{ 1 -  y }  \right)  - i \e} \nn \\
&=&  i  \h \lp  y_{(N-m)}^- - y_{(N-m-1)}^- \rp  
\exp \left[ - i p^+ \left(y_{(N-m-1)}^- - y_{(N-n)}^- \right) \right. \nn \\
\ata \left. \left\{ \left(\sum_{j=0}^{N-m-1}  \frac{x_D^j + (x_T^m)^j }{(1-y)} \right)
+ x_B + x_L^{(N-m-1)} - \left( \sum_{j=1}^m z_j + \frac{(z^m_L)_j}{ 1 -  y }  \right)  - \sum_{j=0}^{N-m-2} x_j \right\} \right] 
\eea
Similarly, the integration over $x^\p_{(N-n-1)}$ requires the contour to be closed at $- i \infty$ and 
thus orders ${y^\p}^-_{(N-n)} > {y^\p}^-_{(N-n-1)} $.

Integrating over single poles such as the above may be continued until $x_{p+1}$ in the 
complex conjugate ($c.c.$) and $x^\p_{q+1}$ in the amplitude.  At this point, there are two double 
poles in the $c.c.$ in $x_p$ and $z_1$, and two double poles in the amplitude in $x_q^\p$ and $z^\p_1$.
Eq.~\ref{O_NN_denom_1}, after all the contour integrations up to $x_p$ and $x^\p_q$ have been performed may 
be expressed as (with integral signs implied), 
\bea
\md^{NN}_{q,p;n,m} \!\!\!\!&=&\!\!\!\! \prod_{i=1}^{N-m} \prod_{k=1}^{N-n} d^3 y_i d^3 y^\p_k 
\frac{ d^2 p_\perp^{i-1} }{ (2\pi)^2 }  \frac{ d^2 {p^\p_\perp}^{k-1} }{ (2\pi)^2 } 
\frac{ dy d^2 l_\perp }{ (2\pi)^3 y }  \frac{ d^2 p_\perp^{(N-m)} }{ (2 \pi)^2 }
\prod_{i=1}^{m} \prod_{j=1}^n d^3 \zeta_i d^3 \zeta^\p_j 
\frac{d z_i d^2 k_\perp^i}{ (2\pi)^3 } \frac{ d {z^\p}_j d^2 {k^\p}_\perp^k }{ ( 2 \pi )^3 }  
\nn \\
%
\ata  d^2 {l_q}_\perp  \kd^2 \left( {l_q}_\perp + l_\perp - \slm_{i=0}^{(N-m)} p^i_\perp - \slm_{j=1}^m k^j_\perp \right) 
\left\{ \prod_{i=0}^p \frac{d x_i}{2\pi}   \frac{ -i }
{ 2q^- \left[   \lp \slm_{j=0}^{i} x_i - x_D^i  \rp  - x_B   - i\e   \right]  }   \right\}  \nn \\
\ata 
\left\{  
\frac{ -i \prod_{i=p+1}^{N-m-1}  \h ( y_{i+1}^-  -  y_i^- )  } 
{ 2q^- (1-y) \left[ \left( \slm_{j=0}^{p} x_j  - \frac{x_D^j + (x_T^m)^j }{(1-y)} \right)
- x_B - x_L^p + \left( \slm_{j=1}^m z_j + \frac{(z^m_L)_j}{ 1 -  y }  \right)  - i \e \right] } 
\right\} \nn \\
\ata   	
\frac{1}{(p^+)^2} 
\left( \frac{1}{2 q^- (1 - y)} \right)^{N-m-p-1} \frac{1}{2 q^- (1-y)} 
\left( \frac{1}{2q^-(1-y)}\right)^{N-n-q-1}  \nn \\
\ata \left\{ \prod_{k=q}^{0} \frac{d x_k }{ 2 \pi }
\frac{ i }{  2q^- \left[   \left(  \slm_{l=0}^{k} x^\p_l - {x^\p}_D^l   \right)  - x_B  + i \e \right]   }  
\right\} \nn \\
\ata \left\{  
\frac{ i \prod_{j=q+1}^{N-n-1} \h ( y_{j+1}^- - y_j^- ) } 
{ 2q^- (1 - y)\left[ \left(  \slm_{l=0}^q  x^\p_l - \frac{ {x_D^\p}^l - ({x^\p}^m_T)^l }{ 1- y }  \right)   
- x_B -  {x^\p}^q_L  +  \left( \slm_{l=1}^n  z^\p_l  + \frac{ {z^\p}^l_L  }{1-y} \right)   + i\e  \right] } 
\right\} \nn \\
\ata \prod_{i=1}^m \frac{-i}{-2q^-y \left[ \slm_{k=i}^m z_k - \frac{(z_L^m)_k}{y} + i\e  \right]} 
\prod_{j=1}^n \frac{i}{-2q^-y \left[ \slm_{l=j}^n z_l - \frac{(z_L^n)_l}{y} - i \e\right]} \nn \\
\ata \prod_{i=0}^{N-m-1} e^{  i p_\perp^i \x ( y_\perp^i  - {y^\p}_\perp^{(N-n)} ) } 
\prod_{ j=1 }^{m} e^{ i k_\perp^j \x ( \zeta_\perp^j - {y^\p}_\perp^{(N-n)} ) } 
\prod_{i=0}^{N-n-1} e^{- i {p^\p}_\perp^i \x ( {y^\p}_\perp^i - {y^\p}_\perp^{(N-n)} )  } 
\prod_{ j=1 }^{n} e^{- i {k^\p}_\perp^j \x ( {\zeta^\p}_\perp^j - {y^\p}_\perp^{(N-n)} )  }  \nn \\
\ata \prod_{i=p+1}^{N-m} e^{-i p^+ y_i^- \left(  \frac{x_D^i + (x_T^m)^i - \kd x_L^i  }{1-y}   \right) } 
\prod_{j=q+1}^{N-n} e^{i p^+ {y^\p}_j^- \left(  \frac{ {x^\p}_D^j + ({x^\p}_T^m)^j - \kd {x^\p}_L^j  }{1-y}   \right) } 
\prod_{i=0}^{p} 
e^{-ip^+ x_i \lp y_i^- - y_{p+1}^- \rp} \prod_{j=0}^q e^{i p^+ {x^\p}_j \lp {y^\p}_j^- - {y^\p}_{q+1}^- \rp } \nn \\
\ata \prod_{i=1}^m e^{  - i p^+ z_i \lp \zeta_i^- - y_{p+1}^- \rp } 
\prod_{j=1}^n e^{ i p^+ {z^\p}_j \left( {\zeta^\p}_j^- - {y^\p}_{q+1}^- \rp  }  
\,\,\, \times e^{ - i p^+ y_{ ( p + 1 ) }^- 
\left[ \left( \slm_{k=0}^{ p }  \frac{ x_D^k + ( x_T^m )^k  }{ ( 1 - y ) } \right)
+ x_B + x_L^p  - \left( \slm_{k=1}^m \frac{ ( z^m_L )_j }{ 1 -  y }  \right)   \right] }  \nn \\
\ata e^{  i p^+ {y^\p}_{ ( q + 1 ) }^- 
\left[ \left( \slm_{l=0}^{ q }  \frac{ {x^\p}_D^l + ( {x^\p}_T^m )^l  }{ ( 1 - y ) } \right)
+ x_B + {x^\p}_L^p  - \left( \slm_{l=1}^n \frac{ ( {z^\p}^n_L )_j }{ 1 -  y }  \right)   \right] }.
\label{O_NN_denom_2} 
\eea
\end{widetext}
The $p_\perp, y_\perp$ are two dimensional vectors with a Euclidean metric. The new momentum fractions, $\kd x_L^i$ and 
$\kd {x^\p}_L^j$ are defined as in Refs.~\cite{Majumder:2007ne}, 
\bea
x_L^i - x_L^{i-1} &=& \frac{ -2 y l_\perp \x \sum_{k=0}^i p_\perp^k + 2 y l_\perp \x \sum_{k=0}^{i-1} p_\perp^{k} }
{2 p^+ q^- y (1 - y)} \nn \\
&=& \frac{ - 2l_\perp \x p_\perp^i}{2 p^+ q^- (1-y)} = - \frac{\kd x_L^i}{1 -y }.
\eea
The momentum fraction on the amplitude side, $\kd {x^\p}_L^j$, has an analogous definition and like $\kd x_L^i$ 
depends only on the transverse momenta of the gluons which strike the propagating quark:
\bea
{x^\p}_L^j - {x^\p}_L^{j-1} &=& \frac{ -2 y l_\perp \x \sum_{l=0}^j {p^\p}_\perp^l + 2 y l_\perp \x \sum_{l=0}^{j-1} {p^\p}_\perp^{l} }
{2 p^+ q^- y (1 - y)} \nn \\
&=& \frac{ - 2l_\perp \x {p^\p}_\perp^j}{2 p^+ q^- (1-y)} = - \frac{\kd {x^\p}_L^j}{1 -y }.
\eea

The first set of non-trivial integrations are those over $x_p$ and $x^\p_q$, which involve two  
poles. It should be pointed out that this is not a true double pole and are indeed two separate 
single poles: one may simply take the sum of the residues at the two poles. 
Isolating the integration over $x_p$, we obtain, 
\begin{widetext}
\bea
\mbx\!\!\!\!&&\int \frac{d x_p}{2 \pi} \frac{ -i e^{ -i x_p p^+ \lt(  y_p^- - y_{ ( p+1 ) }^-  \rt) }  }
{ \lt[  \slm_{i=0}^p  x_i - x_D^i  - x_B - i\e\rt]  } 
\mbx\frac{ -i }
{ \lt[  \lt\{ \slm_{i=0}^p  x_i - \frac{ x_D^i + (x_T^m)^i  }{ 1 - y } \rt\}  
- x_B - x_L^p + \lt\{ \slm_{i=1}^m z_i + \frac{(z_L^m)_i }{ 1 - y } \rt\}  - i\e \rt] } \nn \\
&=& -i \h \lt( y_{ ( p+1 ) }^- - y_p^-  \rt) e^{ ip^+ \slm_{i=0}^{p-1} x_i \lt( y_p^- - y_{(p+1)}^- \rt) } 
e^{- i p^+ \lt( x_B  + \lt\{ \slm_{i=0}^p x_D^i \rt\} \rt) \lt(   y_p^- - y_{(p+1)}^-\rt) }\\
\ata \lt[ 
\frac{  e^{- i p^+ \lt( x_L^p + \lt\{ \slm_{i=0}^p \frac{ yx_D^i + (x_T^m)^i }{ 1 - y } \rt\}  
- \lt\{ \slm_{i=1}^m  z_i + \frac{ ( z_L^m )_i }{ 1 - y } \rt\} \rt) \lt( y_p^- - y_{(p+1)}^- \rt) } }
{ x_L^p + \lt\{ \slm_{i=0}^p \frac{ y x_D^i + (x_T^m)^i }{ 1 - y } \rt\} - 
\lt\{ \slm_{i=1}^m  z_i + \frac{ ( z_L^m )_i }{ 1 - y } \rt\} - i\e }  
- \frac{ 1 }
{ x_L^p + \lt\{ \slm_{i=0}^p \frac{ y x_D^i + (x_T^m)^i }{ 1 - y } \rt\} - 
\lt\{ \slm_{i=1}^m  z_i + \frac{ ( z_L^m )_i }{ 1 - y } \rt\} + i\e   } \rt]. \nn 
\eea

There is a completely analogous expression for the integral over $x^\p_q$ in the amplitude 
which involves the integral over two poles. The remaining integrations over the $x_i$ for $i<p$ and
over $x^\p_j$ with $j<q$ are rather trivial and consist of simple contour integrations as 
performed in Refs.~\cite{Majumder:2007hx,Majumder:2008zg}.
When the set of propagators and associated phase factors in the equation above is combined with 
the propagators and phase factors in the gluon propagators there exist two poles for the momentum
fraction $z_1$ in the complex conjugate and $z_1^\p$ in the amplitude. Isolating the integrand 
of the $z_1$ integration, we obtain, 
\bea
&& \int \frac{ d z_1 }{ 2 \pi }  \frac{i}{ \slm_{i=1}^m z_i - \frac{ (z_L^m)_i }{y} + i\e  }
\lt[ \frac{i  e^{- i p^+ \lt( x_L^p + \lt\{ \slm_{i=0}^p \frac{ yx_D^i + (x_T^m)^i }{ 1 - y } \rt\}  
- \lt\{ \slm_{i=1}^m  z_i + \frac{ ( z_L^m )_i }{ 1 - y } \rt\} \rt) \lt( y_p^- - y_{(p+1)}^- \rt) }
e^{ -ip^+ z_1 \lt( \zeta_1^- - y_{p+1}^- \rt) } }
{\lt\{ \slm_{i=1}^m  z_i + \frac{ ( z_L^m )_i }{ 1 - y } \rt\}
 - x_L^p - \lt\{ \slm_{i=0}^p \frac{ y x_D^i + (x_T^m)^i }{ 1 - y } \rt\} + i\e } \right. \nn \\
&-& \left. \frac{i e^{-ip^{+} z_{1} \lt( \zeta_{1}^{-} - y_{p+1}^{-}  \rt) } }
{ \lt\{ \slm_{i=1}^m  z_i + \frac{ ( z_L^m )_i }{ 1 - y } \rt\} 
-  x_L^p - \lt\{ \slm_{i=0}^p \frac{ y x_D^i + (x_T^m)^i }{ 1 - y } \rt\} - i\e  }  \rt] \nn \\
&=& \h ( \zeta_{1}^{-} - y^{-}_{p} )  e^{- i p^+ \lt( x_L^p + \lt\{ \slm_{i=0}^p \frac{ yx_D^i + (x_T^m)^i }{ 1 - y } \rt\}  
- \lt\{ \slm_{i=2}^m  z_i + \slm_{i=1}^m\frac{ ( z_L^m )_i }{ 1 - y } \rt\} \rt) \lt( y_p^- - y_{(p+1)}^- \rt) } \nn \\
\ata \left\{  \frac{ ie^{- i p^+ \lt( - \slm_{i=2}^m z_i  + \slm_{i=1}^m \frac{(z_L^m)_i}{y}  \rt) \lt( \zeta_1^- - y_p^-\rt) } 
- i e^{- i p^+  \lt( - \slm_{i=2}^m z_i  - \slm_{i=1}^m \frac{(z_L^m)_i}{1-y} + x_L^p + 
\lt\{ \slm_{i=0}^p \frac{ y x_D^i + (x_T^m)^i }{ 1 - y } \rt\}  \rt) \lt( \zeta_1^- - y_p^-\rt)  }     }  
{ \slm_{i=1}^{m} \frac{(z_{L}^{m})_{i}}{y(1-y)}   - x_{L}^{p} 
- \lt\{ \slm_{i=0}^p \frac{ y x_D^i + (x_T^m)^i }{ 1 - y } \rt\}  }   \right\} \nn \\
&-& \left\{  \frac{ \h ( \zeta_1^-  - y_{p+1}^- )   i e^{ -i p^+ 
\left(  - \slm_{i=2}^m z_i  + \slm_{i=1}^m \frac{ (z_L^m)_i}{y} \right) \lt( \zeta_1^- - y_{p+1}^- \rt)  } 
+  \h ( y_{p+1}^- - \zeta_1^-)  i  e^{ -ip^+ \lt(  - \slm_{i=2}^m z_i  - \slm_{i=1}^m \frac{(z_L^m)_i}{1-y} + x_L^p + 
\lt\{ \slm_{i=0}^p \frac{ y x_D^i + (x_T^m)^i }{ 1 - y } \rt\}  \rt) \lt( \zeta_1^- - y_{p+1}^- \rt) } }
{  \slm_{i=1}^{m} \frac{(z_{L}^{m})_{i}}{y(1-y)}   - x_{L}^{p} 
- \lt\{ \slm_{i=0}^p \frac{ y x_D^i + (x_T^m)^i }{ 1 - y } \rt\}  }   \right\}  \nn \\ 
&=& \frac{e^{ ip^+ \lt( \slm_{i=2}^m z_i - 
\slm_{i=1}^m \frac{(z_L^m)_i}{y} \rt) \lt( \zeta_1^- - y_{p+1}^-   \rt)  } }{(-x_L^{pm})}
\lt[ \h(\zeta_1 - y_p^-)  \lt\{ ie^{-ip^+ x_L^{pm} (y_p^- - y_{p+1}^- )  }  
- i e^{ - ip^+ x_L^{pm} (\zeta_1^- - y_{p+1}^-) }  \rt\} \rt. \nn \\
&-& \lt. i \h( \zeta_1^- - y_{p+1}^- )  
- i \h( y_{p+1}^- - \zeta_1^- )  e^{ -ip^+ x_L^{pm} ( \zeta_1^- - y_{p+1}^- )  }  \rt] .
\eea 
Where the aptly named momentum fraction $x_L^{pm}$ is expressed as 
\bea
x_L^{pm} &=& x_L^p + \slm_{i=0}^p \frac{y x_D^i + (x_T^m)_i}{1 - y} - \slm_{i=1}^m \frac{ (z_L^m)_i }{y(1-y)}
= \frac{ \lt| l_\perp - \slm_{i=1}^m k_\perp^i - y \slm_{j=0}^p p_\perp^j \rt|^2 }{ 2 p^+ q^- y (1 - y )} .
\eea
Similarly, the integral on the amplitude side is given as
\bea
&& \frac{e^{ - i p^+ \lt( \slm_{k=2}^n z^\p_k - \frac{({z^\p}_L^n)_k}{y} \rt) 
\lt( {\zeta^\p}_1^- - {y^\p}_{q+1}^- \rt)  } }{{x^\p}_L^{qn}}
\lt[ \h( {\zeta^\p}_1 - {y^\p}_q^-)  \lt\{ ie^{ i p^+ {x^\p}_L^{qn} ( {y^\p}_q^- - {y^\p}_{q+1}^- )  }  
- i e^{  i p^+ {x^\p}_L^{qn} ( {\zeta^\p}_1^- - {y^\p}_{q+1}^-) }  \rt\} \rt. \nn \\
&-& \lt. i \h( {\zeta^\p}_1^- - {y^\p}_{q+1}^- )  
- i \h( {y^\p}_{q+1}^- - {\zeta^\p}_1^- )  e^{ i p^+ {x^\p}_L^{qn} ( {\zeta^\p}_1^- - {y^\p}_{q+1}^- )  }  \rt] ,
\eea
where the momentum fraction ${x^\p}_L^{qn}$ on the amplitude side is given as, 
\bea
{x^\p}_L^{qn} &=& \frac{  \lt| l_\perp - \slm_{l=1}^n {k^\p}_\perp^l 
- y \slm_{k=0}^q {p^\p}_\perp^k \rt|^2  }{ 2 p^+ q^- y (1-y)}.
\eea
The remaining integrations over the variables $z_2$ to $z_m$, $z^\p_s$ to $z^\p_n$, $x_{p-1}$ to $x_0$ and 
$x^\p_{q-1}$ to $x^\p_0$ can now be performed to complete all contour integrations. The resulting expression for 
$\md^{N N}_{q,p;n,m}$ is now completely ordered in the negative light-cone direction, i.e., all the $y^-,{y^{\p}}^-$, 
$\zeta^-, {\zeta^\p}^-$ locations are strongly ordered:
\bea
\md^{NN}_{q,p;n,m} \!\!\!\!&=&\!\!\!\! \prod_{i=1}^{N-m} \prod_{k=1}^{N-n} d^3 y_i d^3 y^\p_k 
\frac{ d^2 p_\perp^{i-1} }{ (2\pi)^2 }  \frac{ d^2 {p^\p_\perp}^{k-1} }{ (2\pi)^2 } 
\frac{ dy d^2 l_\perp }{ (2\pi)^3 y }  \frac{ d^2 p_\perp^{(N-m)} }{ (2 \pi)^2 }
\prod_{i=1}^{m} \prod_{j=1}^n d^3 \zeta_i d^3 \zeta^\p_j 
\frac{ d^2 k_\perp^i}{ (2\pi)^3 } \frac{ d^2 {k^\p}_\perp^k }{ ( 2 \pi )^3 } d^2 {l_q}_\perp  
\nn \\
\ata 
\kd^2 \left( {l_q}_\perp + l_\perp - \slm_{i=0}^{(N-m)} p^i_\perp - \slm_{j=1}^m k^j_\perp \right) 
\lt( \frac{1}{2q^-} \rt)^{p+1} \frac{1}{2q^-(1-y) } \frac{1}{ x_L^{pm} } 
\frac{1}{2p^+} \lt( \frac{1}{2q^- (1 - y)}\rt)^{N-m-p-1}
\!\!\!\!\!\!  \frac{1}{2q^- (1 - y)} \nn \\
\ata \lt( \frac{1}{ 2q^- (1-y)}  \rt)^{N-n-q-1} \frac{1}{2p^+} \frac{1}{ {x^\p}_L^{qn}} 
\frac{1}{2q^- (1-y)} \lt( \frac{1}{2 q^-} \rt)^{q+1} \lt( \frac{1}{2q^-} \rt)^n \lt( \frac{1}{2q^-} \rt)^m \nn \\
\ata \lt[ \h(\zeta_1^- - y_p^-)  \lt\{ e^{-ip^+ x_L^{pm} y_p^-   }  
-  e^{ - ip^+ x_L^{pm} \zeta_1^-  }  \rt\} 
-  \h( \zeta_1^- - y_{p+1}^- )  e^{ - ip^+ x_L^{pm} y_{p+1}^-  }
-  \h( y_{p+1}^- - \zeta_1^- )  e^{ -ip^+ x_L^{pm} \zeta_1^-   }  \rt] \nn \\
\ata \lt[ \h( {\zeta^\p}_1^- - {y^\p}_q^-)  \lt\{ e^{ i p^+ {x^\p}_L^{qn} {y^\p}_q^-   }  
-  e^{  i p^+ {x^\p}_L^{qn}  {\zeta^\p}_1^-  }  \rt\} 
- \h( {\zeta^\p}_1^- - {y^\p}_{q+1}^- )  e^{   i p^+ {x^\p}_L^{qn}  {y^\p}_{q+1}^- } 
- \h( {y^\p}_{q+1}^- - {\zeta^\p}_1^- )  e^{ i p^+ {x^\p}_L^{qn} {\zeta^\p}_1^- }  \rt] \nn \\
\ata e^{-i x_B p^+ y_0^- + p^0_\perp \x y^0_\perp }
\prod_{i=1}^p \h ( y_i^-  - y_{i-1}^-)  e^{- i x_D^i p^+ y_i^- + i p^i_\perp \x y^i_\perp }
\prod_{i=p+1}^{N-m} \h ( y_i^-  - y_{i-1}^-) 
e^{-i p^+ y_i^- \left(  \frac{x_D^i + (x_T^m)^i - \kd x_L^i  }{1-y}   \right)  + i p^i_\perp \x y^i_\perp } \nn \\
\ata \h (\zeta_m^- > \zeta_{m-1}^- > \ldots > \zeta_1^- )
\prod_{i=1}^m e^{-i p^+ \zeta_i^-  \frac{(z_L^m)_i}{y}   + i  k^i_\perp \x \zeta^i_\perp } 
\h ({\zeta^\p}_n^- > {\zeta^\p}_{n-1}^- > \ldots > {\zeta^\p}_1^- )
\prod_{j=1}^n e^{  i p^+ {\zeta^\p}_j^-  \frac{({z^\p}_L^n)_j}{y}   - i  {k^\p}^j_\perp \x {\zeta^\p}^j_\perp } \nn \\
\ata \prod_{j=1}^{q} \h ( {y^\p}_j^-  - {y^\p}_{j-1}^-) e^{ i {x^\p}_D^j p^+ {y^\p}_j^- - i {p^\p}^j_\perp \x {y^\p}^j_\perp }
\prod_{j=q+1}^{N-n} \h ( {y^\p}_j^-  - {y^\p}_{j-1}^-) 
e^{ i p^+ {y^\p}_j^- \left(  \frac{ {x^\p}_D^j + ({x^\p}_T^n)^j - \kd {x^\p}_L^j  }{1-y}   \right)  
- i {p^\p}^j_\perp \x {y^\p}^j_\perp } .
\eea

In the equation above, the exponentials containing the dot products of transverse positions and momenta have 
changed. We have defined a new transverse momentum, 
\bea
\vec{p^\p}^{N-n}_\perp &=& \slm_{i=0}^{N-m} \vp^{\,i}_\perp + \slm_{j=1}^{m} \vk^j_\perp 
- \slm_{i=0}^{N-n-1} {\vec{p^\p}}^i_\perp  - \slm_{j=1}^{n} {\vec{k^\p} }^i_\perp .
\eea
Since there is no integration over ${y^\p}^0_\perp$, we may shift the ${p^\p}^0_\perp$ integration as 
$d^2 {p^\p}^0_\perp \ra d^2 {p^\p}^{N-n}_\perp$. This allows the nesting of the various interactions, 
where transverse momenta from the amplitude and complex conjugate may be combined in pairs. This 
has to be done in a way so as to obtain the maximal enhancement in length and color and will be 
carried out in the next section. 

\end{widetext}


\section{Color factors and length enhancement of Matrix elements}


With the completion of the contour integrals over both light cone components of the incoming momenta and the simplification 
of the numerator Dirac structure, there remains the sum over all interactions and the simplification of the color 
factors. Unlike the case of photon radiation, the color factor in this case is very diagram dependent and depends 
strongly on where the radiated gluons attach to the quark line as well as the number of cross connections between 
the scattering of the gluon and that of the final quark. The determination of the color factor has to be 
carried out in tandem with the estimation of the 
length enhancement achieved by each diagram. As a result, in contrast to the case of photon radiation the first 
step is to decompose the expectation of the $2N$ gluon operators in the nucleon state. This is followed by an 
evaluation of its color factor and length enhancement. Only after this can the sum over emission points 
of the radiated gluon be carried out.

\subsection{Color factor and length enhancement for the fully symmetric diagram}

\begin{figure}[htbp]
\resizebox{3.5in}{3.1in}{\includegraphics[0.5in,0in][12.5in,10.7in]{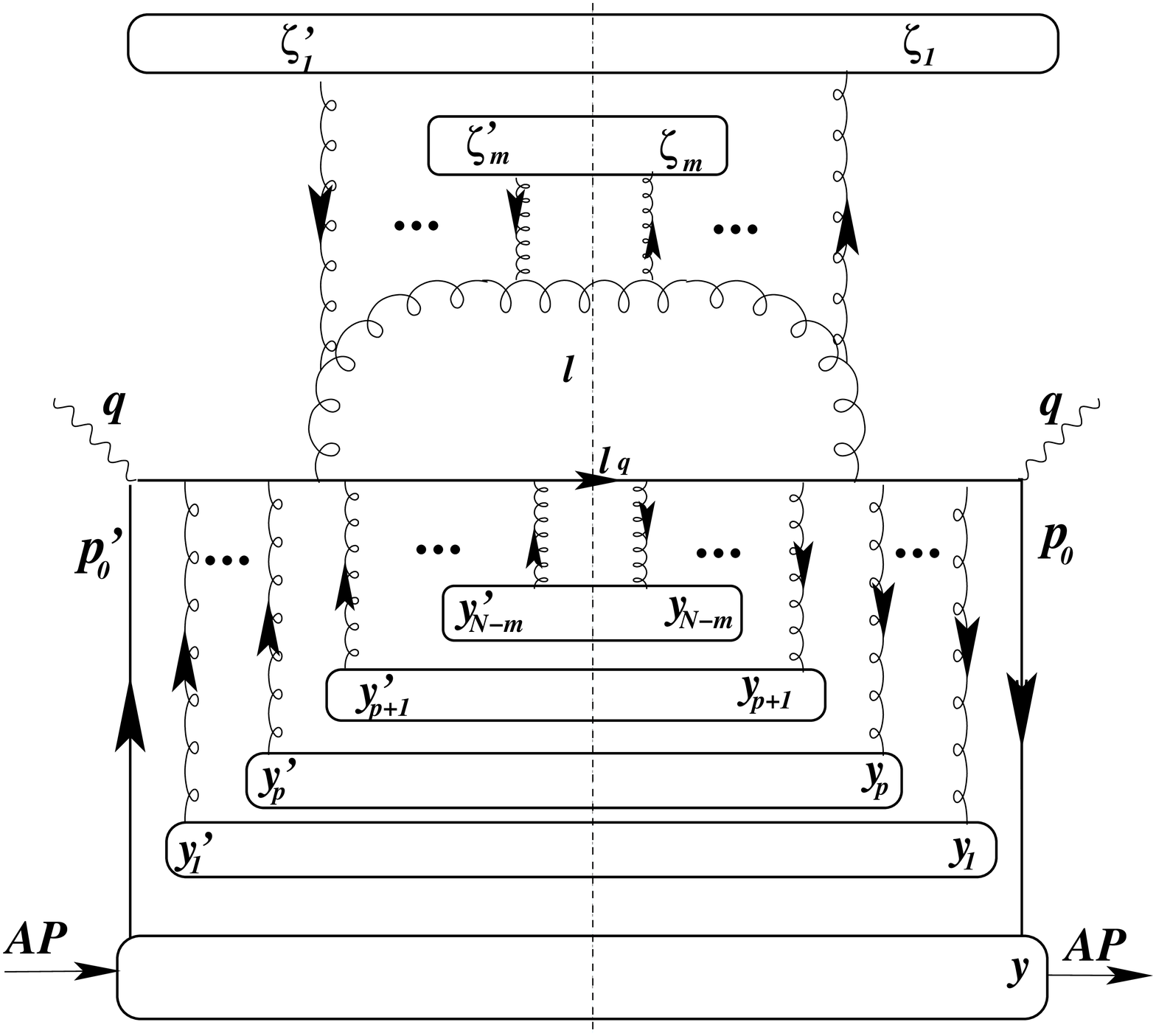}} 
    \caption{ DIS on a nucleus with a produced quark and an emitted gluon scattering $N$ times symmetrically in the 
amplitude and complex conjugate.}
    \label{fig3}
\end{figure}

Given the different color factors for different diagrams we need an 
organization scheme. We begin with the completely symmetric case where $m=n<N$ and $p=q<N$. 
In this case we may nest the various interactions as indicated by the diagram in Fig.~\ref{fig3}. 
Each blob represents one nucleon off which the quark or gluon may scatter. As in the case of transverse 
broadening~\cite{Majumder:2007hx} and photon production~\cite{Majumder:2007ne}, we will ignore 
the expectation of more than two operators per nucleon state. We will not explicitly 
consider the expectation of two successive gluon operators on the same side of the cut line in the 
same nucleon state. These do not induce any change in the transverse momentum distribution and 
constitute unitarity corrections. They are included in the final result by a normalization of the 
resulting transverse momentum distribution~\cite{Majumder:2007hx}.  

The decomposition of the nuclear state as indicated by Fig.~\ref{fig3}, may be carried out as in Refs.~\cite{Majumder:2007hx,Majumder:2007ne}: 
The nucleus is approximated as a weakly interacting homogeneous gas of nucleons. 
Such an approximation is only sensible at very high energy, where, due to 
time dilation, the nucleons appear to travel in straight lines almost independent of each other 
over the interval of the interaction of the hard probe. In a sense, all forms of correlators  
between nucleons (spin, momentum, etc.) are assumed to be rather suppressed. 
As a result, the expectation of 
the $2N$ gluon and 2 quark operators in the nuclear state may be decomposed as 

\bea
&& \lc A;p | \psibar(y^-,y_\perp)\g^+ \psi(0) \prod_{i=1}^{2N} A^{+}_{a_i}(y_i)| A; p\rc \nn \\
&=& A C^A_{p_1}  \lc p_1 | \psibar(y^-,y_\perp)\g^+ \psi(0) \prod_{i=1}^{2N} A^{+}_{a_i}(y_i)  | p_1 \rc
\nn \\
&+& C^A_{p_1,p_2}  \lc p_1| \psibar(y^-,y_\perp)\g^+ \psi(0) | p_1 \rc \nn \\
\ata \lc p_2 | \prod_{i=1}^{2N} A^{+}_{a_i}(y_i)  | p_2 \rc + \ldots , \label{operators_nuclear_state}
\eea
\nt 
where, the factor $C^A_{p_1}$ represents the probability to find a nucleon in the 
vicinity of the location $\vec{y}$, which is a 
number of order unity (it is the probability that one of $A$ nucleons distributed in a 
volume of size $c A$ will be 
found within a nucleon 
size sphere centered at $\vec{y}$). The remaining coefficients $C^A_{p_1,\ldots}$ represent the weak 
position correlations between different nucleons. The overall factor of $A$ arises from 
the determination of the origin (the location $0$ in 
the equation above) in the nucleus, 
which may be situated on any of the $A$ nucleons. 
Solely for the current discussion, we reintroduce the quark operators $\psibar(y^-,y_\perp)$ and $\psi(0)$ in the 
above equation. 

It is clear from the 
above decomposition that the largest contribution arises from the term where the expectation of 
each partonic operator is evaluated in separate nucleon states as the $\vec{y}_i$ integrations 
may be carried out over the nuclear volume. 
As a nucleon is a color singlet, any combination of quark or gluon field strength insertions in 
a nucleon state  must itself  be restricted to a  color singlet combination. As a result, the expectation of 
single partonic operators in nucleon states is vanishing. The first (and hence largest) non-zero contribution 
emanates from the terms where the quark operators in the singlet color combination are evaluated in a 
nucleon state and the $2N$ gluons are divided into pairs of singlet combinations, with each singlet 
pair evaluated in a separate nucleon state. This reduces the discussion to $N+1$ nucleons, one in 
which the original quark is struck by the virtual photon and the remaining $N$ off which the struck 
quark scatters on its way out of the nucleus. As has been demonstrated in Ref.~\cite{Majumder:2008jy}, under the 
assumption that there are no correlations between the nucleons, the combinatorial factor reduces to,
\bea
C^A_{p_1,\ldots,p_{N+1}} = A C^A_{p_1} \left(\frac{ \rho}{2p^+}\right)^N,
\eea
where, $C^A_{p_1}$ varies with the nuclear density profile $\rho$ and is equal to 
unity within the nuclear radius for a hard sphere distribution.

Further simplifications arise in the evaluation of  gluon pairs in a singlet combination in the nucleon 
states by carrying out the $y_\perp$ integrations. The basic object under consideration is  (ignoring the 
longitudinal positions and color indices on the vector potentials)

\bea
\mbx & & \int  d^2 y^i_\perp  d^2 {y^\p}^j_\perp \lc p | A^{+}  (\vec{y}^i_\perp )  A^{+}  ( \vec{y^\p}^j_\perp ) | p \rc \nn \\
&\times & e^{-i x^i p^+ y_i^-}  e^{ i p^i_\perp \x y^i_\perp}   
e^{i {x^\p}^j p^+ {y^\p}_j^-}  e^{ - i {p^\p}^j_\perp \x {y^\p}^j_\perp}  \nn \\
&=& (2\pi)^2 \kd^2( {\vp}^{\,i}_\perp - {\vec{p^\p}} ^j_\perp )  \int d^2 y_\perp  e^{-i x^i p^+ ( y_i^-  -  {y^\p}_j^- )} \nn \\
&\times& 
 e^{ i p_\perp \x y_\perp} \lc p | A^{+} (\vec{y}_\perp/2 )  A^{+}  ( - \vec{y}_\perp/2 ) | p \rc , \label{two_gluon_cor}
 \eea 
where, $y_\perp$ is the transverse gap between the two gluon insertions and $p_\perp = (p^i_\perp + p^j_\perp)/2$. 
The fractions $x_i$ and $x^\p_i$ represent any of the possible types of momentum fractions $x_D, x_T, x_L $
and any combination thereof which appear in the expression for $\Op^{N,N}_{p,p;m,m}$. The physics of the 
above equation is essentially the transverse translation symmetry of the two gluon correlator 
in a very large nucleus. 

One will note that the two dimensional delta function over the transverse 
momenta has removed an integration over the transverse area of the nucleus thus reducing the 
overall $A$ enhancement that may be obtained. This is then used to equate the transverse 
momenta emanating from the two gluon insertions in the amplitude and complex conjugate amplitude. 
This also simplifies the longitudinal phase factors which now depends solely on the 
longitudinal positions of the two gluon insertions. The assumption of short distance 
color correlation length (an assumption valid both in the deconfined plasma as well as in color confined 
cold nuclear matter) leads to the constraint that the longitudinal positions of the gluon operators 
from the amplitude and complex conjugate, whose expectation values are sought in the same nucleon 
state, cannot be more further apart than the color correlation length, i.e.,
\bea
&& \int dy^- d{y^\p}^- \lc p | A^{+} (y^-,\vec{y}_\perp/2 )  A^{+}  ( {y^\p}^-, - \vec{y}_\perp/2 )  | p \rc \nn \\
&\simeq& \int dy^- y_c^- \lc p | A^{+} (y^-,\vec{y}_\perp/2 )  A^{+}  ( y^-, - \vec{y}_\perp/2 )  | p \rc \label{amp_cc_constraint}
\eea
In other words the two locations $y^-$ and ${y^\p}^-$ have been restricted to be in close proximity of each other and
each single nucleon expectation of two gluon operators may introduce at most one factor of length enhancement from 
the remaining $y^-$ integration.

The requirement that the two gluon operators be in the color singlet state, may be enforced by averaging over the 
colors of the two gluon operators, 
\bea
\lc p | A^+_a A^+_b | p \rc = \frac{\kd^{ab}}{ (N_c^2 - 1)} \lc p | A^+_a A^+_a | p \rc. \label{glue_color_conf} 
\eea
A similar constraint on the quark antiquark expection, yields,
\bea
\lc p | \psibar_i\g^+ \psi_j | p \rc = \frac{\kd_{ij}}{ N_c } \lc p |  \psibar \g^+ \psi| p \rc. \label{quark_color_conf}
\eea

The expectation of the quark operators as well as the associated phase factor $ \exp\lt( -i x_B p^+ y_0^- + p^0_\perp \x y^0_\perp \rt)$ and combinatorial factors $A C^A_{p_1}$ may now be extracted and combined to form the nuclear 
structure function of the struck quark. 
Given these simplifications, the full expression for the multiple scattering of the struck quark off $N$ gluons in the symmetric pattern of Fig.~\ref{fig3}, $\Op^{NN}_{p,p;m,m}$, may be written as, 

\begin{widetext}
\bea
\Op^{NN}_{p,p;m,m} \!\!\!\!&=&\!\!\!\! \prod_{i=1}^{N-n} d y_i^-  
\frac{ d^2 p_\perp^i }{ (2\pi)^2 }   
\frac{ dy d^2 l_\perp }{ (2\pi)^3 y } 
\prod_{i=1}^{m} d \zeta^-_i 
\frac{ d^2 k_\perp^i}{ (2\pi)^2 }  d^2 {l_q}_\perp  
\frac{\kd^2 \left( {l_q}_\perp + l_\perp - \slm_{i=0}^{(N-m)} p^i_\perp - \slm_{j=1}^m k^j_\perp \right) P(y)  }
{ \left( \vl_\perp - \slm_{i=1}^m \vk^i_\perp - y \slm_{i=0}^{p} \vp^i_\perp \right)^2  } \nn \\
\ata \lt[ \h(\zeta_1^- - y_p^-)  \lt\{  2  
-  e^{ - ip^+ x_L^{pm} \lt ( \zeta_1^- -  y_p^- \rt)}  - e^{  ip^+ x_L^{pm} \lt ( \zeta_1^-  - y_p^- \rt)} \rt\} 
-  \h( \zeta_1^- - y_{p+1}^- )  
\lt\{ 2 e^{ - ip^+ x_L^{pm} \lt( y_{p+1}^- - y_p^-\rt) }  \rt. \rt. \nn \\
&-& \lt.  2 e^{ - ip^+ x_L^{pm} \lt( y_{p+1}^- - \zeta^-\rt) } \rt\} 
-  \left. \h( y_{p+1}^- - \zeta_1^- ) \h(\zeta_1^- - y_p^-) 
\lt\{ 2 e^{ -ip^+ x_L^{pm} \lt( \zeta_1^- - y_p^- \rt)  } - 2 \rt\} \rt] \nn \\
\ata \lt( \frac{\rho}{2p^+} \rt)^N \prod_{i=1}^p \lt\{ \h ( y_i^-  - y_{i-1}^-)  e^{ i p^i_\perp \x y^i_\perp } 
\frac{\kd^{a_i a^\p_i} \lc p | A^{o_i}  A^{o_i} | p \rc }{N_c^2 - 1}  \rt\}
\prod_{i=p+1}^{N-m} \lt\{ \h ( y_i^-  - y_{i-1}^-) 
e^{ i p^i_\perp \x y^i_\perp }  \rt.
\nn \\
\ata \lt. 
\frac{\kd^{a_i a^\p_i} \lc p | A^{o_i}  A^{o_i} | p \rc }{N_c^2 - 1} \rt\} 
\h (\zeta_m^- > \zeta_{m-1}^- > \ldots > \zeta_1^- ) 
\prod_{i=1}^m e^{ i  k^i_\perp \x \zeta^i_\perp } 
\frac{\kd^{c_i c^\p_i} \lc p | A^{o_i}  A^{o_i} | p \rc }{N_c^2 - 1} \nn \\
\ata \frac{1}{N_c} \tr \lt[ \prod_{i=1}^p t^{a_i} t^{b_0} \prod_{i=p+1}^{N-m} t^{a_i} 
\prod_{i=N-m}^p t^{a^\p_i} t^{b^\p_0} \prod_{i=p}^1 t^{a^\p_i} \rt] \prod_{i=1}^m (f^{b_{i-1} c_i b_i} )
\kd^{b_n b_n^\p} \prod_{i=m}^1 (-f^{ b_i^\p c_i^\p b_{i-1}^\p }) , \label{eqn_for_fig_3}
\eea
\end{widetext}
where, $\lc p | A^{o_i}  A^{o_i} | p \rc$ is short hand for 
\bea
&& \int d \kd y_i e^{- i p^+ x_G^i \kd y_i^-  } \\
\ata \lc p | A^{o_i} (y_i^- + \kd y_i^-/2, y^i_\perp/2) A^{o_i} (y_i^- - \kd y_i^-/2, -y^i_\perp/2)  | p \rc. \nn
\eea
The overall color factor for the Feynman rule above may be evaluated by calculating the trace over the color 
matrices and dividing by the factors obtained from averaging over colors, which yields,
\bea
C^{N,N}_{p,p;m,m} &=& \frac{ \lt( C_F \rt)^{(N-m+1)} C_A^{m}  }{ N_c (N_c^2 - 1)^{N} } \nn \\ 
&=& \frac{ \lt( \frac{ (N_c^2 - 1) }{2N_c} \rt)^{(N-m+1)} N_c^{m}  }{ N_c (N_c^2 - 1)^{N} } \nn \\ 
&=& \frac{N_c^2 - 1}{2N_c} \lt( \frac{1}{2 N_c} \rt)^{N-m} \lt( \frac{N_c}{N_c^2 - 1} \rt)^{m} .
\eea

Under the assumption that the parton propagates through a long extended medium, matrix elements are  
enhanced by the length traversed through the medium. Without any constraints except those of 
Eq.~\eqref{amp_cc_constraint} 
which connect the scattering locations on the amplitude with those on the complex 
conjugate, the maximal length enhancement factor is 
\bea
\lt( \int_0^L d y^- \rt)^N = L^N . 
\eea
One can isolate the leading length enhanced contribution for a given configuration 
by analyzing the various $\h$-function constraints that limit the length integrations. For the diagram of 
Fig.~\ref{fig3} corresponding to Eq.~\eqref{eqn_for_fig_3}, we note that all the scattering points on the 
quark and the gluon are ordered in the ($-$)-light-cone coordinate i.e., 
$\zeta_m^- > \ldots > \zeta_1^-$, $y_{N-m}^- > \ldots > y_0^-$. 
In addition the gluon scattering points for the first term in square 
brackets are forced to be ahead of $y_p^-$ on the quark, i.e., $\zeta_1^- > y_p^-$ . For the second
term in square brackets $\zeta_1^- > y_{p+1}^-$, while for the third term $y_{p+1}^- > \zeta_1^- > y_p^-$.

The largest length factor for each of the terms may be estimated by performing all integrations over the entire length 
with a unit integrand. For a single \emph{stream} of constrained 
integrals, this procedure yields the well known result,
\bea
\int_0^L d y_0 \int_{y_0}^L d y_1 \ldots \int_{y_{N+1}}^L d y_N = \frac{ \lt( \int_0^L d y \rt)^N }{N!} .
\eea
For the case of two separate streams joined at a point $y_p$ as in the case the first term in the square brackets of Eq.~\eqref{eqn_for_fig_3}, this procedure yields the length enhancement factor, 
\bea
L^{N,N}_{p,p;m,m;1} = \frac{  (N-p)!  \lt( \int_0^L d y \rt)^N  }{ (N-m-p)! (m)! N!  }.
\eea
To make it easier to analyze these factors we make the replacement, $m_q=N-m-p$ ($m_q$ is the number of times the 
quark scatters after radiating the gluon), which yields,
\bea
L^{N,N}_{p,p;m,m;1} = \frac{  (m+m_q)!  \lt( \int_0^L d y \rt)^N  }{ m_q! m! N!  }.
\eea
Similarly, the length enhancement factor for the second term in square brackets is given as,
\bea
L^{N,N}_{p,p;m,m;2} &=& \frac{ (N-p-1)! \lt( \int_0^L d y \rt)^N  }{ (N-m-p-1)! (m)! N!} \nn \\
&=&  \frac{ (m+m_q-1)! \lt( \int_0^L d y \rt)^N  }{ (m_q-1)! (m)! N!}.
\eea
Finally the length enhancement factor for the third factor with the square brackets of Eq.~\eqref{eqn_for_fig_3} may 
be obtained by noting that $\zeta_1^-$ is locked between $y_p^-$ and $y_{p+1}^-$, thus this is similar to the 
the first length enhancement factor but with the replacements $m \ra m-1$, $p \ra p+1$, i.e.,
\bea
L^{N,N}_{p,p;m,m;3} &=& \frac{ (N-p-1)! \lt( \int_0^L d y \rt)^N  }{ (N-m-p)! (m-1)! N!} \nn \\
&=& \frac{ (m+m_q-1)! \lt( \int_0^L d y \rt)^N  }{ (m_q)! (m-1)! N! }.
\eea

Inspection of these terms suggests that the largest contributions arise from the case where $m+m_q \ra N$, i.e., 
most of the scattering happens after the gluon has been radiated. We can now complete the analysis of this 
term by expanding the transverse  dependence of the hard part. In this case, the hard part consists of everything 
except the matrix elements and the associated phase factors which contain the transverse momentum controlled 
by the matrix elements. The first step is to institute the condition that $l_\perp \gg k^i_\perp$. This simplifies the 
momentum fractions $x_L^{pm}$ which appear in the phase factors to $x_L = l_\perp^2/(2 p^+ q^- y (1-y))$. 
One can also expand the sum of transverse momenta in the denominator as, 
\bea
&& \frac{1}{ \lt( \vl_\perp  - \sum_{i=1}^n \vk^i_\perp - y \sum_{i=1}^p \vp^i_\perp \rt)} \nn \\
&=& \frac{1}{l_\perp^2}  +  \frac{ 2 \vl_\perp \x \lt( \sum_{i=1}^n \vk^i_\perp + y \sum_{i=1}^p \vp^i_\perp \rt)}
{l_\perp^4} \nn \\
&-& \frac{ \lt( \sum_{i=1}^n \vk^i_\perp + y \sum_{i=1}^p \vp^i_\perp \rt)^2}
{l_\perp^4} - \ldots 
\eea

Since contributions are dominated by smaller values of $p$, 
as a first approximation, one may neglect the contribution with 
large $p$ and focus on the $p=0$ limit. Note that in the case of 
multiple emissions leading to the change of a fragmentation function, 
the momentum fractions carried out by the gluon is restricted to a 
small fraction of unity. This consideration adds additional weight to the neglect of these
terms, especially for the case of  leading hadron suppression calculations. 
We may further shift the ${l_q}_\perp$ integration as ${l_q}_\perp = {l_q}_\perp - l_\perp$, 
to represent the difference in the transverse momentum of the radiated gluon and the 
final quark.
Incorporating the above mentioned simplifications, 
expanding the transverse momentum dependent term above and the 
transverse momentum dependent $\kd$-function yields, 
\bea
\prod\limits_{i,j=1}^{m,n}&&\!\!\!\!\!\!\!\!\hf \frac{\prt^2}{\prt^2 k^i_\perp }\frac{\prt^2}
{\prt^2 p^j_\perp } \left[ \lt\{ \frac{ 1  }{l_\perp^2}    
+ \frac{ 2 \vl_\perp \x \lt(  \slm_{l=1}^m \vk^l_\perp \rt)  
- \lt( \slm_{l=1}^m \vk^l_\perp \rt)^2}{l_\perp^4} \right. \right. \nn \\
\mbox{}\!\!\!\!\!\!\!\!\!\!\!\!+&& \!\!\!\!\!\!\!
\left. \frac{  \lt(2 \vl_\perp \x \slm_{l=1}^m \vk_\perp^l \rt)^2 }{l_\perp^6} \right\} \\
\mbox{}\!\!\!\!\!\!\!\!\!\!\!\! \times && \!\!\!\!\!\!\! 
\lt. \kd^2 \lt( \vec{l_q}_\perp - \slm_{l=1}^m \vk^l_\perp -\slm_{l=1}^n \vp^l_\perp \rt)
\rt]_{\vk^i_\perp=\vp^j_\perp = 0}\!\!\!\!\!\!\!\! | \vk^i_\perp|^2 | \vp^j_\perp|^2 
\nn \\	
\mbox{}\!\!\!\!\!\!\!\!\!\!\!\! = && \!\!\!\!\!\!\! \lt(\hf\rt)^{N} \lt[ \frac{   
\lt( \nabla_{{l_q}_\perp}^2  \rt)^N \kd^2 \lt( \vec{l_q}_\perp  \rt)  }{l_\perp^2  } \rt. \nn \\
\mbox{}\!\!\!\!\!\!\!\!\!\!\!\!- && \!\!\!\!\!\! 
\frac{2 m \vl_\perp \x \nabla_{{l_q}_\perp}  
\lt( \nabla_{{l_q}_\perp}^2  \rt)^{N-1}  \kd^2 \lt( \vec{l_q}_\perp  \rt) }{l_\perp^4} \nn \\
\mbox{}\!\!\!\!\!\!\!\!\!\!\!\! + && \!\!\!\!\!\!  \lt. \frac{4 m 
\lt( \nabla_{{l_q}_\perp}^2  \rt)^{N-1}   \kd^2 \lt(  \vec{l_q}_\perp  \rt) }{l_\perp^4} \rt]
\!\!\prod\limits_{i,j=1}^{m,m_q} |\vk^i_\perp|^2 | \vp^j_\perp|^2 . \nn
\eea

The factors of incoming transverse momentum, both on the gluon line ($|k^i_\perp|^2$) as well 
as those on the quark line ($|p^j_\perp|^2$) may be combined with the position 
dependent matrix elements to yield the Lorentz force correlator obtained in Ref.~\cite{Majumder:2007hx} (suppressing color indices and longitudinal momentum and position 
arguments): 
\bea 
&& \!\!\!\!\!\!\!\!\!\!\!\int d^2y^i_\perp |\vk^i_\perp|^2 e^{i \vk^i_\perp \x \vec{y^i}_\perp} 
\lc p | A^+ (y_\perp/2) A^+ (-y_\perp/2) |p\rc \\
&=& \int d^2y^i_\perp e^{i \vk^i_\perp \x \vec{y^i}_\perp} 
\hf \lc p | F^{+,\A_i} (y_\perp/2) F_{\A_i ,}^+ (-y_\perp/2) |p\rc .\nn
\eea
In the equation above, $F^{+,\A_i}$ represent the gluon field strength operators. 
It should be noted that $\A_{i}$ only runs over the transverse components. 
In the full expression for Eq.~\eqref{fig3}, the incoming transverse momenta $k_\perp^i$, 
$p_\perp^j$ appear only in the exponentials and may be integrated out to confine the 
transverse locations to the origin. Hence we define the longitudinal position dependent 
coefficient, 
\bea
\bar{D}(y_i^-) &=&  2 \pi^2 \A_s \int \frac{d \kd y_i^-}{2\pi} \int \frac{d^2k_\perp}{ (2\pi)^2} d^2 y^i_\perp 
e^{i \vk^i_\perp \x \vec{y^i}_\perp }  \\
\ata \lc p | F^{+ \mu}(y_i^- + \kd y_i^-,y^i_\perp/2) F_\mu^+(y_i^-,-y^i_\perp/2) |p \rc  \nn \\
&=& 2 \pi^2 \A_s \int \frac{d \kd y_i^-}{2\pi} \lc p |  F^{+ \mu}(y_i^- + \kd y_i^-) F_\mu^+(\kd y_i^-) |p \rc .  \nn
\eea
As the integration over the transverse momentum $k_\perp^i$ restricts the transverse separation $y^i_\perp$ to zero, 
this is no longer explicitly written out in the final expression.

This last simplification allows one to express the cross section for $N$ symmetric scattering with no scattering 
prior to emission (i.e., restricting to only the first $\h$-function of Eq.~\eqref{eqn_for_fig_3} with $p\ra 0$) as, 
\begin{widetext}
\bea
\Op^{N,N}_{0,0;m,m} &=& \int \frac{dy d^2 l_\perp}{(2\pi)^3} d^2 l_q \frac{ 4 \pi \A_s C_F P(y)}{l_\perp^2 y}
\lt[ \frac{   
\lt( \nabla_{{l_q}_\perp}^2  \rt)^N \kd^2 \lt( \vec{l_q}_\perp  \rt)}{ N!} \frac{N!C_A^m C_F^{N-m}}{m! (N-m)!}  
 \lt( \frac{ \rho \int_0^{L^-} d y_i^- \bar{D} (y_i^-) }{2p^+ ( N_c^2 - 1 )}\rt)^{N-1} \right. \nn \\
&+&  \lt\{ - \frac{2 \vl_\perp \x \nabla_{{l_q}_\perp}  
\lt( \nabla_{{l_q}_\perp}^2  \rt)^{N-1}  \kd^2 \lt( \vec{l_q}_\perp  \rt) }{(N-1)!l_\perp^2} 
+ \frac{4 
\lt( \nabla_{{l_q}_\perp}^2  \rt)^{N-1}   \kd^2 \lt(  \vec{l_q}_\perp  \rt) }{(N-1)!l_\perp^2} \rt\} \nn \\
\ata\lt.\frac{(N-1)!}{(m-1)! (N-m)!}  C_A^m C_F^{N-m} \lt( 
\frac{ \rho \int_0^L dy^- \bar{D} (y^-) }{2 p^+ (N_c^2 - 1)} \rt)^{N-1}  \rt] \nn \\
\ata \rho \int d \zeta_1^-  \frac{ \bar{D} (\zeta_1^-) }{2p^+ (N_c^2 - 1)}  \lt\{  2  
-  2 \cos \lt(p^+ x_L \zeta_1^- \rt) \rt\} \label{symmetric_simple}
\eea
\end{widetext}
Summing over $m$ for fixed $N$ combines the gluon and quark color factors, as
\bea
\sum_{m=0}^N \frac{N!}{m! (N-m)!}  C_A^m C_F^{N-m} = (C_A + C_F)^N.
\eea
As in the case of photon production we absorb the color, phase and density factors into the 
definition of two new diffusion coefficients:
\bea 
D &=& \frac{(C_F + C_A)\rho \bar{D}}{(2p^+ (N_c^2 - 1))},  \label{joint_diff_coeff} \\
E(x_L)^{\pm} &=& (C_F + C_A)\rho \int d \zeta_1^-  \frac{ \bar{D} (\zeta_1^-) }{2p^+ (N_c^2 - 1)} 
e^{ \pm ip^+ x_L \zeta_1^- } .\nn
\eea

Summing over $N$ allows us to resum the multiple derivatives of the two dimensional delta function as~\cite{sneddon}, 
\bea
&&\sum_{N=0}^\infty \frac{ \lt(\int d y^- D (y^-) \rt)^N \lt( \nabla_{{l_q}_\perp}^2  \rt)^N}{N!} 
\kd^2 ({l_q}_\perp )  \nn \\
&=& \frac{e^{- \frac{{l_q}_\perp^2}{ 4 \int d y^- {D} (y^-) } } }{4 \pi \int d y^- {D} (y^-)}. \label{diff_eqn_soln}
\eea
Using the above identity one may easily resum Eq.~\eqref{symmetric_simple} over all scatterings. The final result is 
a very simple extension of the result of next-to-leading twist.

\subsection{Color factor and length enhancement for more complicated diagrams}

\begin{figure}[htbp]
\resizebox{2.5in}{2.1in}{\includegraphics[0.5in,0in][12.5in,10.7in]{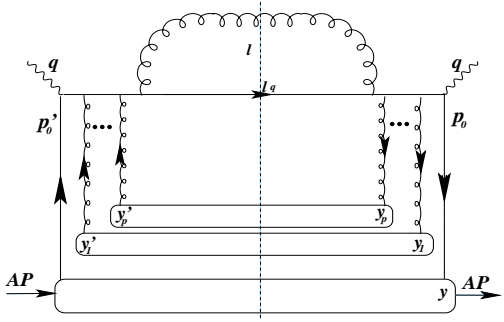}} 
    \caption{Simplest example of scattering prior to emission.}
    \label{fig4}
\end{figure}

After demonstrating the outcome of a rather simple and leading case, the color and length factors of 
more complicated diagrams will be outlined in this subsection. The entire set of diagrams may be divided 
into six different types with different color and length enhancement factors. Any diagram may be 
constructed as a combination of these six types of diagrams. The first and simplest type of 
scattering are those which occur before the gluon is radiated, as shown in Fig.~\ref{fig4}. 
All scatterings on the amplitude and 
complex conjugate are correlated in pairs and ordered in time. The general color factor for $p$ scatterings is 
given as, 
\bea
&& \tr\lt[ \prod_{i=1}^m t^{a_i^\p} t^{b_0^\p} \hat{C} t^{a_0} \prod_{i=m}^1 t^{a_i} \rt] 
\frac{1}{N_c} \prod_{i=1}^m \frac{\kd^{a_i a^\p_i}}{N_c^2 -1} \nn \\
&=& \frac{C_F^m}{N_c (N_c^2 - 1)^m} \tr \lt[  t^{b_0^\p} \hat{C} t^{a_0} \rt].
\eea
As all the scatterings are correlated and ordered the diagram has the obvious length enhancement factor of 
$(y_p)^p/p!$ where we have assumed that the quark starts at the origin and the $p^{th}$ scattering 
occurs at the location $y_p$. The $N_c \times N_c$ dimensional operator $\hat{C}$ represents further scattering 
encountered by the quark and gluon pair after emission.
This particular type of scattering has no color entanglement effects with other types of scattering.

\begin{figure}[htbp]
\resizebox{2.1in}{2.1in}{\includegraphics[0.5in,0in][12.5in,10.7in]{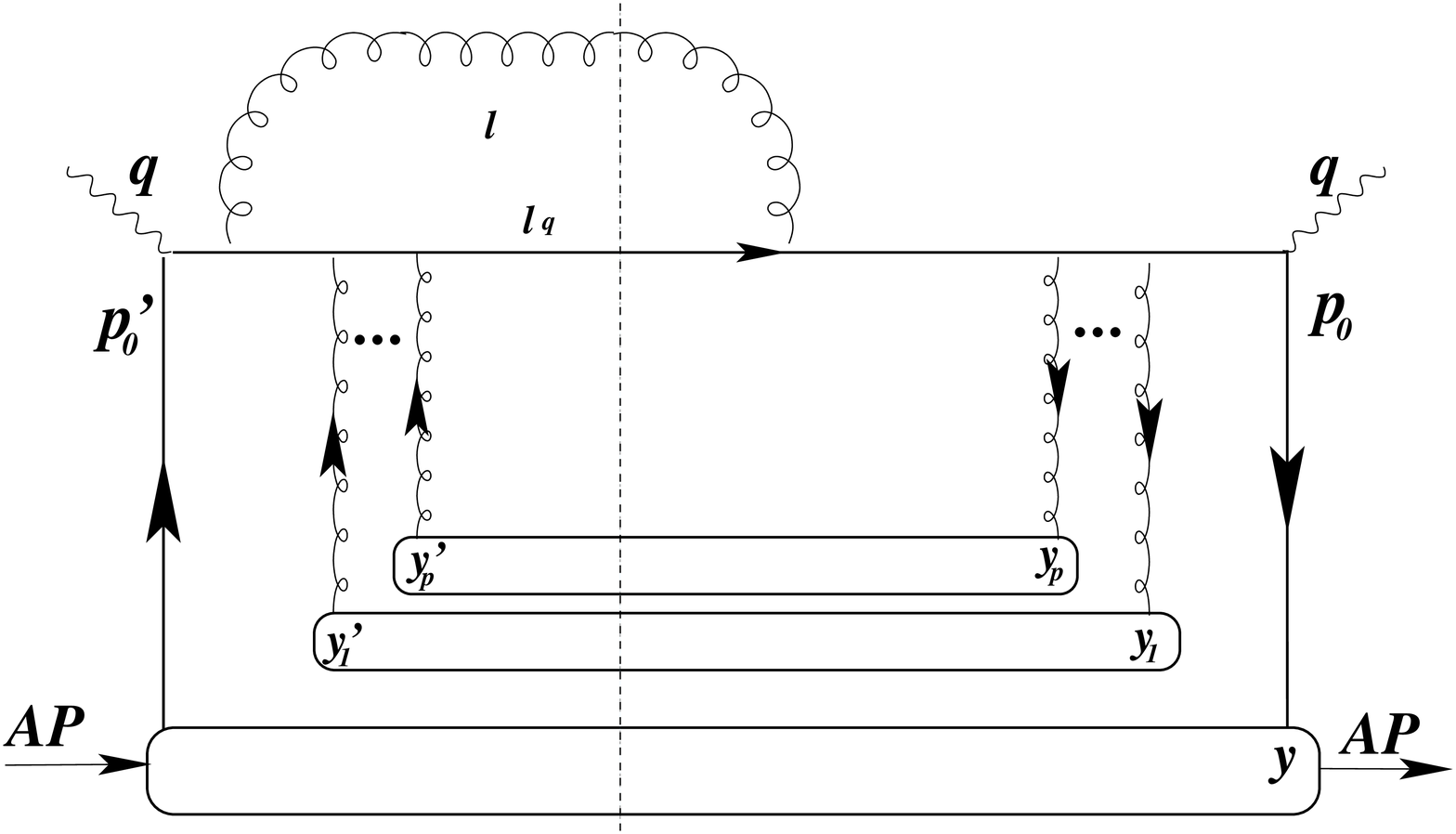}} 
    \caption{An interference effect where scattering of the quark before the emission is correlated 
with scattering of the quark after the emission.}
    \label{fig5}
\end{figure}

\begin{figure}[htbp]
\resizebox{2.1in}{2.1in}{\includegraphics[0.5in,0in][12.5in,10.7in]{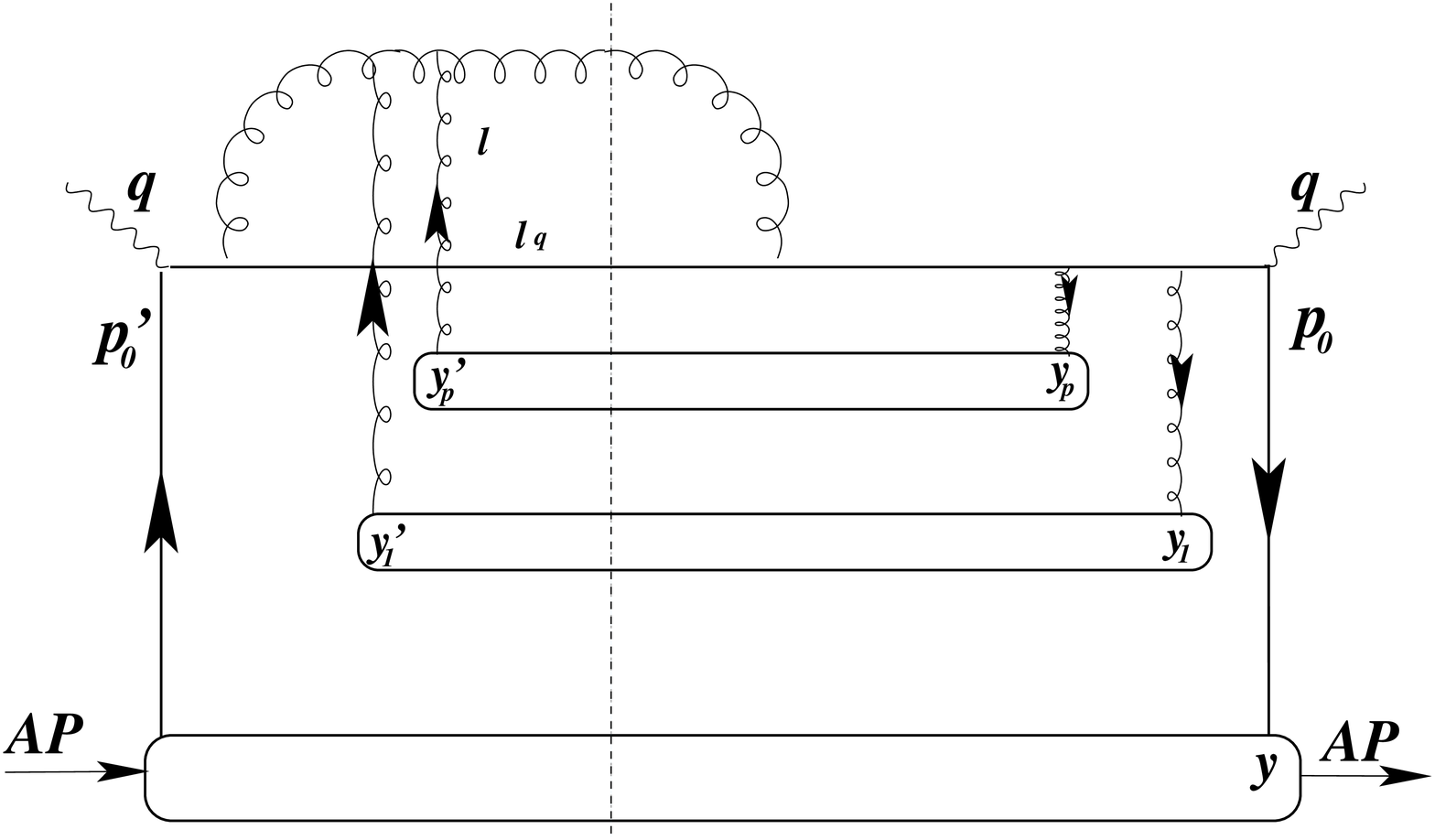}} 
    \caption{An interference effect where the scattering of the quark before emission is correlated with the 
scattering of the gluon after emission.}
    \label{fig6}
\end{figure}

The next type of scattering diagrams are sketched in Fig.~\ref{fig5} and involve the interference of 
scattering that occurs before the emission of the gluon correlated with scattering on the quark produced after the 
emission. The color factor for this type of diagram is given as 
\bea
&& \tr\lt[ \prod_{i=1}^m t^{a_i^\p} t^{b_0^\p} \hat{C}  \prod_{i=m}^1 t^{a_i} t^{a_0} \rt] 
\frac{1}{N_c} \prod_{i=1}^m \frac{\kd^{a_i a^\p_i}}{N_c^2 -1} \nn \\
&=&  \frac{ \lt( C_F - C_A/2 \rt)^m}{N_c (N_c^2 - 1)^m} \tr \lt[  t^{b_0^\p} \hat{C} t^{a_0} \rt].
\eea
Very similar to this type of diagram is the case where the scattering on the quark before emission are 
correlated with scattering on the gluon after emission. The color factor for this type of diagram, shown 
in Fig.~\ref{fig6} is given as, 
\bea
&& \tr\lt[ \prod_{i=1}^m (-it^{a_i^\p}) t^{b_0^\p} \hat{C} t^{a_0} \rt] 
\frac{1}{N_c} \prod_{i=1}^m f^{b_{i-1} a_i b_i }  \kd^{b_0 a_0} \hat{C}_2^{b^m b_0^\p}\nn \\
\ata \prod_{i=1}^m \frac{\kd^{a_i a^\p_i}}{N_c^2 -1} 
=  \frac{ \lt( C_A/2 \rt)^m \tr \lt[  t^{b_m} t^{b_0^\p} \hat{C} \rt] }{N_c (N_c^2 - 1)^m} \hat{C}_2^{b^m b_0^\p}.
\eea
Since these scatterings are also ordered and correlated, we find that the length enhancement factor in this case for $m$ 
scattering is simply $(y_{m+p} - y_p)^m/m!$ where $y_p$ is the location where the scattering begins and $y_{p+m}$ is the 
location of the last scattering of this type. The color contributions from such diagrams are completely untangled with 
the case of Figs.~\ref{fig4} and \ref{fig5} and also with later scattering on the quark and gluon after emission. The 
operators $\hat{C}$ and $\hat{C}_2^{b^m b_0^\p}$ indicate further scattering. 

The remaining three types of scattering are all in the final state. Two of these are rather simple and
involve correlated ordered scattering of the final quark or the gluon which yields the obvious 
multiplicative color factors of $C_F^n/(N_c^2 -1)^n$ or 
$C_A^n/(N_c^2 -1)^n$ and have no interference with each other yielding separate length enhancement factors of 
$(y_f - y_{p+m})^n/n!$. As these are final state effects, $y_f$ is the extent of the medium and 
$y_{p+m}$ is the location of the last interference term between scattering in the initial and final states.
As these color and length factors are rather trivial no diagram corresponding to these is presented. 

\begin{figure}[htbp]
\resizebox{2.5in}{2.1in}{\includegraphics[0.5in,0in][12.5in,10.7in]{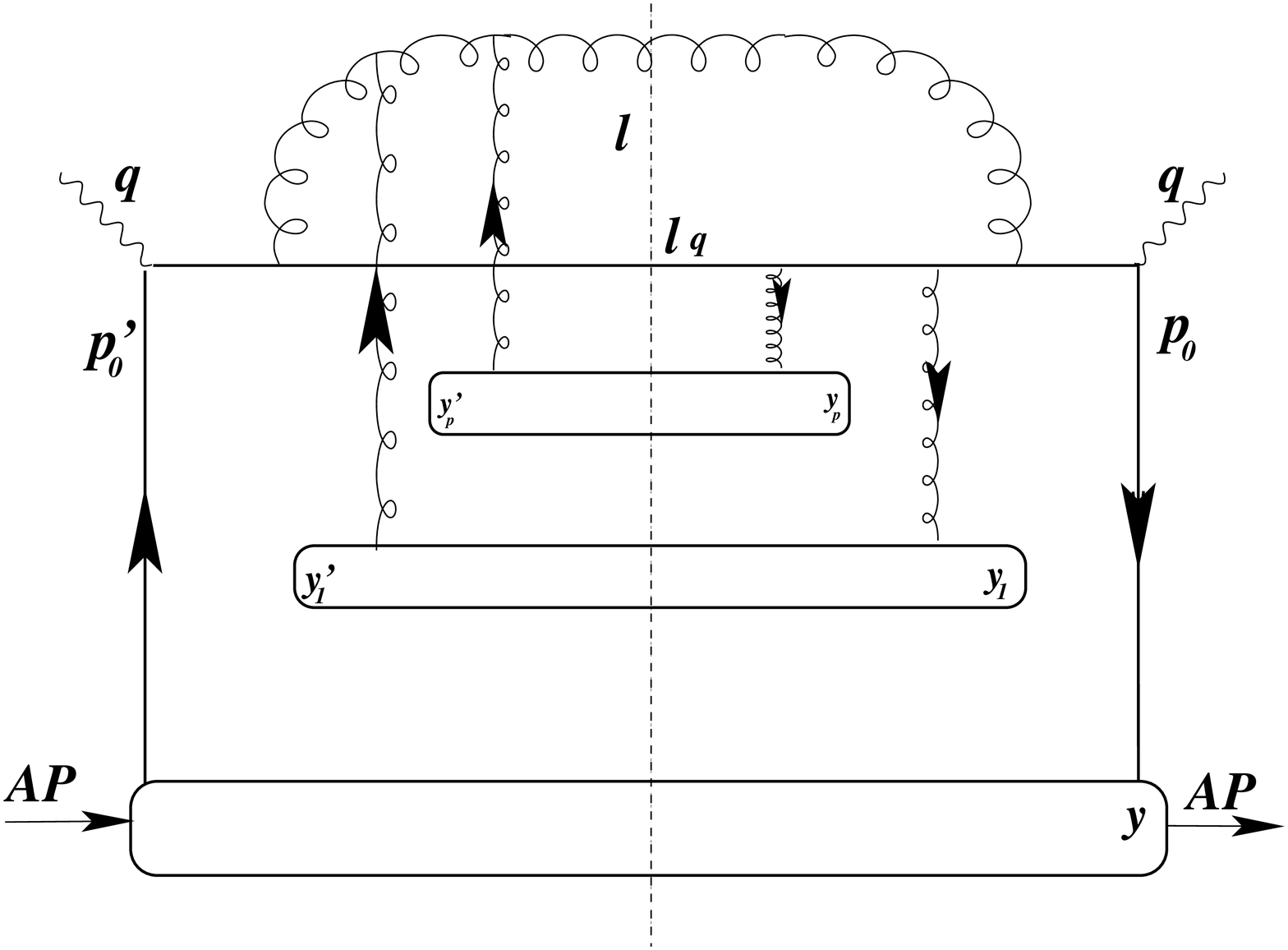}} 
    \caption{An interference effect where the scattering of the the quark after emission is correlated with the 
scattering of the gluon after emission.}
    \label{fig7}
\end{figure}

The most complicated situation, which is ignored by most other formalisms is the interference between 
quark scattering in the final state and gluon scattering in the final state along with entangled 
diagrams where such scattering interchanges with ordered correlated scattering on the final quark and 
gluon. Analyzing the purely interfering case of the type of diagrams shown in Fig.~\ref{fig7} with 
$n$ scatterings, the length enhancement factor may be immediately surmised to be of the 
form $(y_f - y_{p+m})^n/n!$. Note that these contributions and those of scattering purely on the 
final quark and the gluon may become entangled. However at the same order of coupling, the former 
will always yield a larger length enhancement factor in the limit of multiple scattering. 

The color factor for diagrams of the type shown in Fig.~\ref{fig7}, may be calculated as, 
\bea
C_n = \tr \lt[ t^{c_n} \prod_{i=1}^n (-it^{a_i^\p}) t^{c^0}  \rt] \frac{1}{N_c} \prod_{j=0}^n f^{c_{j-1} a_j c_j} \frac{ \kd^{a_j a^\p_j} }{ N_c^2 - 1}.
\eea
There is no general formula for terms of this type. For the case of $n=1$, we obtain, 
$C_1 = \frac{C_F iC_A/2}{N_c^2 - 1}$; for $n=2$ we obtain $C_2=0$. For $n=3$ we obtain, $C_3 = \frac{iN_c}{N_c (N_c^2 - 1)^2}$. 
Further scatterings cannot be done analytically and numerical methods yield that these color factors are 
very suppressed compared to all other types of scattering at the same order of coupling. These will henceforth
be ignored in the remaining analysis. Diagrams with interchanging attachments where  the first scattering on a 
quark in the amplitude interferes with a gluon in the complex conjugate followed by a scattering on the gluon in 
the amplitude interfering with a scattering on the quark in the complex conjugate will yield similar color factors.
There remain the case of entangled diagrams, which may be shown to be suppressed as well and will also be ignored.


\section{ The multiple scattering Resummed single gluon emission kernel. }


The analysis of the previous sections will now be incorporated to obtain the 
leading result for the all-twist resummed single gluon emission kernel. 
At this juncture we will also attempt to connect with other leading formalisms 
for jet energy loss. In particular we will discuss two limits of the calculation 
which deal with the results of Refs.~\cite{AMY} (referred to as the Arnold-Moore-Yaffe (AMY) formalism) 
and those of Refs.~\cite{ASW} (referred to as the Armesto-Salgado-Wiedemann (ASW) formalism).

We begin this final section with a discussion of two types of diagrams which have not 
yet been discussed. Virtual contributions for which the cut line only goes through 
one side of the radiated gluon without cutting the radiated gluon or the final quark. 
These virtual corrections are important for inclusive observables and 
tend to unitarize the single gluon emission kernel. In single hadron inclusive~\cite{Majumder:2009zu} 
(and also for parts of multi-hadron inclusive calculations~\cite{Majumder:2008jy}) 
these diagrams are incorporated by 
insisting the product of the final fragmentation function and 
the integral over the integrated splitting function be subtracted from the real contribution. 
In the case of calculations of the energy lost in a single 
gluon emission or the change in the partonic distributions, which are more exclusive i.e., depend on the 
momentum carried out by both quark and gluon, these diagrams play no role as one is 
more interested in solely the real contributions.

The other type of diagrams are those which have an unequal number of scattering on 
either side of the cut in either real of virtual diagrams. Such diagrams unitarize the 
effect of multiple scatterings. As shown in Ref.~\cite{Majumder:2007hx}, summing over 
all types of multiple scattering and integrating over 
the transverse momentum of a given line should yield a factor of order unity i.e., no 
length enhancement. This effect is introduced into the calculation by seeking normalized 
solutions of the diffusion equation that results from the multiple scattering expansion.


\subsection{The limit of only gluon scattering and leading twist}


The is essentially the limit of large $N_c$, where we assume that only the gluon 
scatters. As demonstrated by the color factors derived in the previous section, the difference 
between a quark and a gluon scattering in the medium induces a factor of 
$C_A/C_F = N_c^2/[2 (N_c^2 - 1)]$ 
per scattering, which is enhanced $n$ times for $n$ scattering. Thus in the limit of multiple 
scattering, the naive expection is that such diagrams dominate over those where the quark also 
scatters. This is not entirely accurate as for a given order the length enhancement factor
for the case of equal scattering on the final quark and the gluon is length enhanced over 
the case of scattering solely on the quark or the gluon. The actual situation is 
intermediate between these two cases. In the small $y$ limit, the scattering on the 
initial quark is suppressed by $y$. 
Also for a computation of the energy loss, the scattering on 
the final quark does not yield any change in 
the momentum dependent part of the expression for the energy loss. This is because, no factor of the 
transverse momentum imparted to the final quark appears in any of the terms of the 
integrand except for the transverse momentum $\kd$-function. The scattering on the 
final outgoing quark however does introduce a non-trivial change in the phase 
factor which will be considered in the next subsection.

One may easily obtain the expressions corresponding to the limit of only gluon scattering 
by setting $m=N$ in Eq.~\eqref{symmetric_simple}. All factors of $C_F$ which arise 
from the scattering of the quark disappear. The sole factor of $C_F$ corresponds 
to the gluon emission vertex. As a result, the only type of transport coefficient that 
plays a role is the gluon transport coefficient $\hat{q}$, i.e., 
the diffusion coefficient of Eq.~\eqref{joint_diff_coeff} simplifies to 
\bea
D = \frac{C_A \rho \bar{D}}{2 p^+ (N_c^2 - 1)}.
\eea
The transverse broadening of a parton which has progressed a distance $L^-$ , is given as
\bea
l_\perp^2 = 4 \int_0^{L^-} d \zeta^- D(\zeta^-)  
=  \int_0^{L^-} d \zeta^- \frac{\hat{q}}{2} 
= \int_0^L d \zeta \hat{q}.
\eea
The factor of 1/2 along with $\hat{q}$ is due to our notation of light cone variables 
where $\zeta^- = 2 \zeta$ and the definition of $\hat{q}$ as $l_\perp^2/L$. Thus we obtain, 
\bea
\hat{q} = 8 D = \frac{8 \pi^2 \A_s C_A  }{N_c^2 - 1} \rho x_T G(x_T)
\eea

The $N^{\rm th}$ order transverse momentum derivatives on the transverse momentum 
$\kd$-function may now be resummed. To be able to resum the result, one has to 
extract the first length integration over $\zeta_1^-$ and order the subsequent 
scatterings. As a result, we obtain,  
\begin{widetext}
\bea
&& \sum_{N=1}^\infty \Op^{N,N}_{0,0;m,m} = 
\sum_{N=1}^\infty \int\frac{dy d^2l_\perp d^2 {l_q}_\perp }{2\pi^2} \frac{\A_s C_F P(y)}{l_\perp^2 y}
\int_0^{L^-} d \zeta^-  D (\zeta^-) 
\lt\{  2  -  2 \cos \lt(p^+ x_L \zeta^- \rt) \rt\} \label{ASW_limit_1}\\ 
\ata \lt[ \frac{   
\lt( \nabla_{{l_q}_\perp}^2  \rt)^N \kd^2 \lt( \vec{l_q}_\perp  \rt)}{ (N-1)!} 
+ N \lt\{ - \frac{2 \vl_\perp \x \nabla_{{l_q}_\perp}  
\lt( \nabla_{{l_q}_\perp}^2  \rt)^{N-1}  \kd^2 \lt( \vec{l_q}_\perp  \rt) }{(N-1)!l_\perp^2} 
+ \frac{4 \lt( \nabla_{{l_q}_\perp}^2  \rt)^{N-1} \kd^2 \lt(  \vec{l_q}_\perp  \rt) }
{(N-1)!l_\perp^2} \rt\}  \rt] 
\lt( \int_{\zeta^-}^{L^-} \!\!\!\!\!\!dy^- D (y^-) \rt)^{N-1} \!\!\!\!\!. \nn
\eea
In order to sum over $N$, and indeed even in the derivation of the expression 
above for a fixed $N$, we have ignored terms which are suppressed by powers of $N$ 
in the large $N$ limit. 
While this allows us to resum the series above into a closed form, it has the 
disadvantage of not reducing to the known limit at $N =1$~\cite{HT}. Summing 
over $N$ we obtain the single gluon emission kernel in the ASW limit of 
multiple scattering as, 
\bea
\sum_{N=1}^\infty \Op^{N,N}_{0,0;m,m}\!\!\! &=& 
\!\!\!\int \frac{dy d^2l_\perp d^2 {l_q}_\perp }{2\pi^2} \frac{\A_s C_F P(y)}{l_\perp^2 y}
\int_0^{L^-} \!\!\!\!\!\!\!d \zeta^-  D (\zeta^-)  
\lt\{  2  -  2 \cos \lt(p^+ x_L \zeta^- \rt) \rt\} 
\lt[  \left(\frac{4 - 2 \vl_\perp \x \nabla_{{l_q}_\perp} }{l_\perp^2} \right) 
\frac{e^{- \frac{{l_q}_\perp^2}{ 4 \int d y^- {D} (y^-) } } }{4 \pi \int d y^- {D} (y^-)} \rt. \nn \\
&+& \lt. \nabla_{{l_q}_\perp}^2   
\frac{e^{- \frac{{l_q}_\perp^2}{ 4 \int d y^- {D} (y^-) } } }{4 \pi \int d y^- {D} (y^-)}
- \lt( \int_{\zeta^-}^{L^-} \!\!\!\!\!\!dy^- D (y^-) \rt) 
\lt\{ \frac{2 \vl_\perp \x \nabla_{{l_q}_\perp}  
\nabla_{{l_q}_\perp}^2 -  4 \nabla_{{l_q}_\perp}^2  }{l_\perp^2} \rt\}
\frac{e^{- \frac{{l_q}_\perp^2}{ 4 \int d y^- {D} (y^-) } } }{4 \pi \int d y^- {D} (y^-)}
 \rt].  \label{ASW_limit_2}
\eea
\end{widetext}
The above expression represents the leading contribution at all twist to the double 
differential distribution of a hard quark and a gluon produced from a hard initial 
quark. While we keep referring to this as the ASW limit, it should be pointed out that we are 
evaluating this in the range $l_\perp^2 \gg k_\perp^2$ to $ l_\perp^2 \gtrsim k^2_\perp$, 
whereas in the original ASW papers, the evaluation is carried out toward the lower limit. 

Guided by our assumption that $l_\perp$ is large, we have expanded the complete 
result of Eq.~\eqref{eqn_for_fig_3} as a series in $k_\perp^2 / l_\perp^2$ 
keeping terms up to $1/l_\perp^4$. Expanding further will lead to more terms. 
However, the Eq.~\eqref{ASW_limit_2} has a special property when applied to inclusive 
observables such as energy loss. To derive the single gluon emission kernel from the 
double differential distribution above, we integrate over ${\vec{l_q}}_\perp$ and over 
the angle of gluon emission. One will immediately note the identity:
\bea
\int d^2 {l_q}_\perp  \frac{e^{- \frac{{l_q}_\perp^2}{ 4 \int d y^- {D} (y^-) } } }{4 \pi \int d y^- {D} (y^-)} = 1
\eea
as well as the vanishing contributions, 
\bea
&&\int d^2 {l_q}_\perp  {\nabla_{l_q}}_\perp^2 \frac{e^{- \frac{{l_q}_\perp^2}{ 4 \int d y^- {D} (y^-) } } }{4 \pi \int d y^- {D} (y^-)} = 0 , \label{gaussian_rules}  \\ 
&\&& \int d^2 {l_q}_\perp  l_\perp\x{\nabla_{l_q}}_\perp \frac{e^{- \frac{{l_q}_\perp^2}{ 4 \int d y^- {D} (y^-) } } }{4 \pi \int d y^- {D} (y^-)} = 0. \nn
\eea
The second identity is true only in the case of unpolarized observables. 
Integrating over ${l_q}_\perp$, we obtain the simple and well know result for single scattering induced emission~\cite{HT}, 
\bea
\sum_{N=1}^\infty \Op^{N,N}_{0,0;m,m} &=& 
\int \frac{dy dl_\perp^2 }{2\pi} \frac{\A_s C_F P(y)}{l_\perp^2 y} \label{mult_scat_resum_to_single_scat}\\
\ata \int_0^{L^-} d \zeta^-  4 D (\zeta^-)  
\lt[  2  -  2 \cos \lt(p^+ x_L \zeta^- \rt) \rt]. \nn
\eea

While the above result may come as a surprise, the primary reason for the cancellation of 
the large number of terms is the unitarity of the 
transverse momentum distribution due to multiple scattering. The sum over diagrams, 
where the scatterings are not nested as in Fig.~\ref{fig3}, restores unitarity to the 
distribution and thus enforces the choice of a normalized Gaussian solution to the 
diffusion equation~[Eq.~\eqref{diff_eqn_soln}]. 
In obtaining the above result we ignored higher transverse momentum 
derivatives which contained larger powers of $l_\perp^2$ in the denominator. Including 
these will introduce corrections to the above simple form. This is tantamount 
to restricting the calculation to the next-to-leading twist part of the all twist 
calculation. As would have been expected this yields the same result as the 
small $y$ limit of the single scattering calculation~\cite{HT}.


\subsection{The full calculation and comparison with the AMY  and ASW schemes}


In this subsection, we incorporate the effect of multiple scattering on both the quark and 
the gluon in the calculation of the single gluon emission kernel. In the preceding section, 
the various types of scattering were divided into six categories. Of these, two types which 
involve the interference between the scattering of the gluon and the quark after emission 
have very small color factors and do not possess larger length enhancements compared to other 
diagrams. These will be ignored in this effort. This leaves only four different types of 
diagrams that need to be considered: scattering on the quark before emission, interference 
between scattering of the quark or gluon after emission with scattering before emission and 
separate scattering of the produced quark and gluon. Diagrams which entangle these contributions 
tend to be suppressed in length factors. Thus these four contributions can be considered 
one after the other separately.

In order to incorporate all these effects. We divide the length of the medium traversed from 
$y_0^-$ to $L^-$ into four regions. The first part from $y_0^-$ to $y_Q^-$ within which only 
the initial quark scatters and the gluon emission has not occurred in either amplitude or 
complex conjugate (Fig.\ref{fig4}). From $y_C>y_Q$ to $y_E$ where there is interference between the scattering 
of the quark in the amplitude (complex conjugate) with scattering of the quark or gluon in the 
complex conjugate (amplitude) [Fig.~\ref{fig5} or Fig.~\ref{fig6}]. 
These are referred to as the cross terms (hence indicated as $y_C$) and continue till the later 
gluon emission at $y_E$. 
We also keep track of the location $\zeta_C$ which is the 
first scattering of the gluon in the complex conjugate (amplitude). The last region is from 
$y_I>y_E$ to $L^-$ which represents the part where the produced quark and gluon scatter 
independently. We also keep track of $\zeta_I \geq y_E$ which is the first scattering of 
the gluon in this regime. Note that $\zeta_C \leq \zeta_I \leq L^-$.  The gluon emission kernel 
will depend on all these locations.

Unlike in the preceding subsection, where only one diffusion coefficient was required as there was 
only one type of scattering (that on the final state gluons), there are four different types of 
scattering in this subsection. As a result, four different types of diffusion coefficients will be
required which differ in their overall color factor only. 
We denominate these as $D_{GG}$ (or simply $D_G$) where both scatterings in the amplitude and complex-conjugate 
occur on the gluon,
\bea
D_{GG} = \frac{C_A \rho \bar{D}}{2 p^+ (N_c^2 - 1)}.
\eea
The standard transport coefficient in jet-quenching $\hat{q} = 8 D_{GG}$.
The case where one scattering is on the gluon and one on the quark is denoted as $D_{GQ}$ or $D_{QG}$ 
interchangeably, and given as
\bea
D_{GQ} = \frac{(C_A/2) \rho \bar{D}}{2 p^+ (N_c^2 - 1)}.
\eea
The case where both scatterings are on the quark and one occurs prior to gluon emission and one 
occurs after will be designated as $D_{Q\bar{Q}}$ and has the expression,
\bea
D_{Q\bar{Q}} = \frac{(C_F - C_A/2) \rho \bar{D}}{2 p^+ (N_c^2 - 1)}.
\eea
It may come as a surprise that this diffusion coefficient is actually negative. 
Multiple such insertions of interfering terms where the quark in the amplitude 
scatters before the emission and the quark in the complex conjugate scatters 
after the emission lead to a shrinking of the transverse momentum between the 
radiated gluon and the final outgoing  quark. The final diffusion coefficient is the 
case where the the quark scatters in both amplitude and complex conjugate and 
both happen either before or after the gluon emission. The diffusion coefficient in
this case is designated as $D_{QQ}$, or simply $D_{Q}$, and given as 
\bea
D_{QQ} = \frac{(C_F) \rho \bar{D}}{2 p^+ (N_c^2 - 1)}.
\eea

The derivation of the full double differential distribution of the cross section as a function of the 
transverse momentum of the quark and the gluon is rather complicated and will be presented 
elsewhere. In this article, we will simply derive an expression for the energy loss distribution. 
From the analysis of the preceding sub-section on the ASW limit we note that in the differential 
distribution, all terms which involve  extra derivatives of the delta function vanish in the integrated 
expression. Also terms which involve one integral to be kept separate from the rest due to a phase 
factor depending on this location will primarily be involved in the higher twist contributions. Thus terms 
with more than one phase factor are suppressed by larger powers of $l_\perp^2$. Thus to obtain the leading 
contribution only the terms with one phase factor need to be retained.

In the interest of simplicity we will not derive the extra contribution from diagrams where the quark scatters 
before the emission without interfering with any other type of scattering. These will be 
included separately at the end of the section. These diagrams were shown 
to have smaller length enhancement factors. They proportional to $y^2$ and are thus 
very suppressed in the small $y$ limit. As a result, they play little role in the computation 
of the medium modified fragmentation function~\cite{Zhang:2003yn} where the range of the gluon momentum 
fraction is limited by $0 < y < 1-z$. The fraction $z$ is the momentum fraction of the detected 
hadron with respect to the initial hard quark's momentum 
and is thus also the minimum momentum fraction that must be carried by the final quark to be able to 
fragment and produce such a hadron.  These diagrams also do not contain a vacuum like contribution in 
either amplitude or complex conjugate and arise as a purely medium dependent piece.  
As a result, the neglect of these terms focuses this part of the calculation to manifestly involve the vacuum contribution 
in either the amplitude or complex conjugate. 
As a result, this part of the calculation is somewhat orthogonal to the calculations in the formalism of 
Arnold, Moore and Yaffe (AMY) where none of the terms contain a vacuum contribution. 
This part of the calculation, 
however has a considerable overlap with the calculation of Armesto Salgado and Wiedemann (ASW) 
except that in the ASW case there is no Taylor expansion in the exchanged momentum i.e., there is 
no explicit assumption $l_\perp \gg k_\perp$.

Invoking the above mentioned simplifications, 
we write the expression for $\Op^{r+s+m+n,q+m+n}_{0,q;s+m,m}$ where 
$q=r+s$. This represents the operator expression for the case where the quark emits a gluon at the origin in the 
amplitude and after $q$ scatterings in the complex conjugate. The $q$ scatterings interfere with $r$ scatterings 
on the produced quark and $s$ scatterings on the gluon in the amplitude. After these interfering terms, 
the produced quark and gluon scatter $m$ and $n$ times in both amplitude and complex conjugate. The full expression 
is written as, 
\begin{widetext}
\bea
\Op^{q+m+n,q+m+n}_{0,q;s+m,m} \!\!\!\!\!&=& 
\!\!\!\!\!\int\!\!\frac{dy d^2l_\perp d^2 {l_q}_\perp }{2\pi^2} \A_s C_F P(y)
C_A^m C_F^n (C_F - C_A/2)^r (C_A/2)^s 
\kd^2 \!\!\lt( {l_q}_\perp - \sum_{i=1}^s k^i_\perp - \sum_{j=1}^r p^j_\perp  
- \sum_{l=1}^m k^l_\perp - \sum_{k=1}^n p^k_\perp \rt)  \nn \\
\ata \frac{ l_\perp -  \slm_{i=1}^s k^i_\perp  - \slm_{l=1}^m k^l_\perp  }
{ \lt( l_\perp -  \slm_{i=1}^s k^i_\perp -  \slm_{l=1}^m k^l_\perp\rt)^2 }
\x \frac{ l_\perp -  y\slm_{i=1}^s k^i_\perp -  \slm_{l=1}^m k^l_\perp  }
{ \lt( l_\perp - y \slm_{i=1}^s k^i_\perp -  \slm_{l=1}^m k^l_\perp\rt)^2 }  \nn \\
\ata \prod_{i=1}^{q} \int d y_i^- \frac{\int d^{3} \kd y_{i} \rho \lc p | A^{+}(y_i^-+\kd y_{i}^{-},0)  A^{+}(y_{i}^{-},-\kd y_{\perp}^{i}) | p \rc }
{2 p^{+}(N_c^2 - 1)}  e^{i k_{\perp}^{i} \kd y_{\perp}^{i}}
\nn \\
\ata 
\prod_{j=1}^{n}  \int d y_j^- \frac{ \int d^{3} \kd y_{j} \rho \lc p | A^{+}(y_j^- + \kd y_{j}^{-},0)   A^{+}( y_{j}^{-}, -\kd y_{\perp}^{j} ) | p \rc }
{ 2p^{+}(N_c^2 - 1)} e^{i k_{\perp}^{j} \kd y_{\perp}^{j} } \nn \\
\ata \prod_{l=1}^{m} 
\int d\zeta_l^- \frac{ \int d^{3} \kd \zeta_{l} \lc p | A^{+}(\zeta_l^- + \kd \zeta_{l}^{-} , 0 )   A^{+} ( \zeta_{l}^{-} ,- \kd \zeta_{\perp}^{l} ) | p \rc} 
{(N_c^2 - 1)}  e^{i k_{\perp}^{l} \kd \zeta_{\perp}^{l}} \nn \\
\ata \lt[ \h(\zeta_I^- - y_E^-)  \lt\{ e^{-ip^+ x_L y_E^-   }  
-  e^{ - ip^+ x_L \zeta_I^-  }  \rt\} 
-  \h( \zeta_I^- - y_I^- )  e^{ - ip^+ x_L y_I^-  }
-  \h( y_I^- - \zeta_I^- )  e^{ -ip^+ x_L \zeta_I^-   }  \rt] \nn \\
\ata \lt[ \h( \zeta_C^- - y_0^- )  \lt\{ e^{ i p^+ x_L y_0^-   }  
-  e^{  i p^+ x_L  \zeta_C^-  }  \rt\} 
- \h( \zeta_C^- - y_C^- )  e^{ i p^+ x_L  y_C^- } 
- \h( y_C^- - \zeta_C^- )  e^{ i p^+ x_L \zeta_C^- }  \rt].  \label{all_twist_non_amy}
\eea
%
As mentioned above, we will consider the case where the initial state scattering on the 
quark is ignored thus we set $y_0^- = 0$. The least power suppressed 
contribution in this limit is the term containing the $e^{ip^+ x_L y_0^-} = 1$ factor.

The second line in Eq.~\eqref{all_twist_non_amy} may be simplified as, 
\bea
&& \frac{1}{l_\perp^2}  - \frac{ (1 - y + y^2) \lt( \slm_{i=1}^s k_\perp^i \rt)^2 }{ l_\perp^4  }  
- \frac{ \lt( \slm_{i=1}^m k_\perp^i \rt)^2 }{ l_\perp^4}
+ 2 (1+y^2) \frac{ \lt( l_\perp \x  \slm_{i=1}^s  k^i_\perp \rt)^2 }{ l_\perp^6   } + 
4 \frac{ \lt( l_\perp \x \slm_{i=1}^m  k^i_\perp \rt)^2  }{ l_\perp^6}.
\eea
In the equation above, we have only retained only terms which are suppressed by at most $l_\perp^4$.  This is all that 
is required for the case of terms containing the $e^{ip^+ x_L y_0^-} = 1$ factor. Terms containing two 
separate phases will require the products  $ ( \sum_{i=1}^s k_\perp^i )^2 ( \sum_{i=1}^m k_\perp^i )^2$ which 
appear in the above expansion suppressed by $l_\perp^6$. While it is not difficult to include these terms and these 
will be included in future numerical calculations, we refrain from including them here as we illustrate the leading 
behavior of the multiple scattering gluon radiation spectrum. 

Expanding the $\kd$-function and the (non-phase) momentum dependent part 
as a series in transverse momentum, we obtain, 
\bea
\Op^{q+m+n,q+m+n}_{0,q;s+m,m} \!\!\!\!\!&=& 
\!\!\!\!\!\int\!\!\frac{dy d^2l_\perp d^2 {l_q}_\perp }{2\pi^2} \A_s C_F P(y)
C_A^m C_F^n (C_F - C_A/2)^r (C_A/2)^s  \lt[  \frac{ \lt( \nabla_{{l_q}_\perp}^2 \rt)^{q+m+n} } {l_\perp^2}
+ \frac{ 4 ( s y + m)  \lt( \nabla_{{l_q}_\perp}^2 \rt)^{q+m+n-1} }{ l_\perp^4 } \rt]  \nn \\
\ata \kd^{2} ( l_{q_{\perp}} ) \prod_{i=1}^{q} \int d y_i^- \frac{ \rho \bar{D}(y_i^-)}{2 p^+ (N_c^2 - 1)}  
\prod_{j=1}^{n}  \int d y_j^- \frac{ \rho \bar{D}(y_j^-)}{2 p^+ (N_c^2 - 1)} 
\prod_{l=1}^{m} \int d\zeta_l^- \frac{ \rho \bar{D}(\zeta_l^-)}{2 p^+ (N_c^2 - 1)} \label{all_twist_non_amy_2}\\
\ata \lt[ \h(\zeta_I^- - y_E^-)  \lt\{ e^{-ip^+ x_L y_E^-   }  
-  e^{ - ip^+ x_L \zeta_I^-  }  \rt\} 
-  \h( \zeta_I^- - y_I^- )  e^{ - ip^+ x_L y_I^-  }
-  \h( y_I^- - \zeta_I^- )  e^{ -ip^+ x_L \zeta_I^-   }  \rt] \nn 
\eea
\end{widetext}

In the equation above, the $q$ integrals run from $y_{0} = 0$ to $\zeta_{I}$ or $y_{I}$ 
which ever comes first. The $q$ integrals are all ordered and 
$r$ of them attach to a quark on both sides of the cut. For terms without the $exp[-ip^{+} x_{L}y_{E}^{-}]$,
 this yields the length enhancement factor of, 
\bea
&& \prod_{i=1}^{q} \int d y_{i}^{-} \h (y_{i}^{-} - y_{i-1}^{-})  \frac{(r+s)!}{r! s!} \nn \\
&=& \frac{\lt( \int_{y_{0}}^{min \lt[y_{I},\zeta_{I}\rt] } d y \rt)^{r}}{r!} 
\frac{\lt( \int_{y_{0}}^{min \lt[y_{I},\zeta_{I} \rt] } d y \rt)^{s}}{s!} 
\eea
For the term containing the $y_{E}^{-}$ phase factor, 
if $y_{E}$ scatters on a quark on both sides, then we take the $r^{\rm th}$ integration out from the product. 
If $y_{E}$ scatters on a gluon then we take the $s^{\rm th}$ integration out from the ordered product, yielding 
the factor, 
\bea
\mbox{}\!\!\!\!\!\!
&& \int_{0}^{min[y_{I},\zeta_{I}]} d y_{E}^{-} \frac{ \rho \bar{D}(y_E^-)}{2 p^+ (N_c^2 - 1)}  e^{-ip^{+} x_{L}y_{E}^{-}}  \\
\mbox{}\!\!\!\!\!\!
\ata\!\!\!\! \lt[  \frac{ \lt( \int_{0}^{y_{E}^{-} } d y \rt)^{r-1} }{(r-1)!}  \frac{\lt( \int_{0}^{ y_{E}^{-} } d y \rt)^{s}}{s!} 
+ \frac{ \lt( \int_{0}^{y_{E}^{-} } d y \rt)^{r} }{(r)!}  \frac{\lt( \int_{0}^{ y_{E}^{-} } d y \rt)^{s-1}}{(s-1)!}  \rt]. \nn
\eea

As a result, including all the length dependent terms, associated with the $y_{E}^{-}$ phase factor, we get, 
\bea
\mbox{}\!\!\!\!\!\!
&& \int_{0}^{L^{-}} d y_{E}^{-}  \frac{ \rho \bar{D}(y_E^-)}{2 p^+ (N_c^2 - 1)}  e^{-ip^{+} x_{L}y_{E}^{-}} \nn \\
\mbox{}\!\!\!\!\!\!
\ata\!\!\!\!
 \lt[  \frac{ \lt( \int_{0}^{y_{E}^{-} } d y \rt)^{r-1} }{(r-1)!}  \frac{\lt( \int_{0}^{ y_{E}^{-} } d y \rt)^{s}}{s!} 
+ \frac{ \lt( \int_{0}^{y_{E}^{-} } d y \rt)^{r} }{(r)!}  \frac{\lt( \int_{0}^{ y_{E}^{-} } d y \rt)^{s-1}}{(s-1)!}  \rt] \nn \\
\mbox{}\!\!\!\!\!\!
\ata\!\!\!\!
 \frac{ \lt( \int_{y_{E}^{-}}^{L^{-}} d\zeta \rt)^{m}  }{ m!  } \frac{ \lt(  \int_{y_{E}^{-}}^{L^{-}}  d \zeta  \rt)^{n} }{ n!}.
\eea
When combined with the transverse momentum derivatives one notes that there is no contribution from the terms 
containing $r-1$ (as these will always contain an extra $\nabla_{{l_q}_\perp}^2$ which is not being summed) 
i.e., where $y^-_E$ represents the scattering on a quark. Summing over $r,s,m,n$,
and integrating over $l_{q_{\perp}}$,  the only surviving term is, 
\bea
\mbx\!\!\!\!\!\!\!\Op(1) \!\!\!&=& 
\!\!\!\!\!\int\!\!\frac{dy d^2l_\perp }{2\pi^2} \A_s C_F P(y)
 \nn \\
\ata \!\!\! \int_{0}^{L^{-}}  \!\!\!\!d y_{E}^{-}  4y D_{GQ} (y^{-}_{E})  e^{- i p^{+} x_{L} y_{E}^{-} } .
\label{final_res_1}
\eea
Note that this term is explicitly $y$ dependent and thus subleading as $y \ra 0$. 

Similarly for terms with the $e^{-i p^{+} x_{L} \zeta_{I}^{-}}$ phase factor, the length dependent terms, after summation over 
all insertions and integration over ${l_{q}}_{\perp}$, yields, 
\bea
\Op(2) &=&  - \int\!\!\frac{dy d^2l_\perp }{2\pi^2} \A_{s} C_F P(y)
 \nn \\ 
 \ata 2 \int_{0}^{L^{-}} \!\!\!\!\!\!d \zeta^{-} 4 D_G (\zeta^{-}) e^{-ip^{+} x_{L} \zeta^{-}} .
\label{final_res_2}
\eea
The two equations above, represent terms which are not squares of amplitudes but rather interference terms. 
Thus we also need to include the complex conjugates. The result of summing the two equations above with 
their complex conjugates may be easily obtained by converting $e^{- i p^{+} x_{L} y_{E}^{-} } $ to $2\cos( p^{+} x_{L} y_{E}^{-} )$
in Eq.\eqref{final_res_1} and $e^{- i p^{+} x_{L} \zeta^{-} } $ to $2\cos( p^{+} x_{L} \zeta^{-} )$ in Eq~\eqref{final_res_2}.

In the results above, we have only obtained terms which contain a cosine of the usual 
phase argument $(x_L p^+ \zeta^-)$ but not the terms proportional to unity. The combination of these 
yields the LPM effect. To obtain the terms which don't contain the argument $x_L p^+ \zeta^-$, we 
have to consider terms with at least two of the sums restricted, i.e., terms where $r=0,s=0$ or terms 
with $m=0,n=0$. 
Considering
these latter contributions changes the overall phase factor, e.g., when $r=0,s=0$ one will have to replace
$y_C^-$ with $y_I^-$ and $\zeta_C^-$ with $\zeta_I^-$ in Eq.~\eqref{all_twist_non_amy}. While this
limit provides the zeroth order contributions to the various diffusion terms that go with 
the $\cos( p^{+} x_{L} \zeta^{-} )$, it also introduces new contributions, i.e., the terms which 
don't depend on the cosine, 
which have to be included separately. This we carry out in the following.

The reader will note that all the sums over the number of scatterings start from a minimum of $1$, e.g., in the first line of Eq.~\eqref{all_twist_non_amy}. This requires that all types of scatterings considered have to have 
taken place at least once. We also need to include the cases where a particular type of scattering did not 
take place. The simplest situation is the case where $m=n=r=s=0$. In this case there will not be any of the 
phase factors or in-medium matrix elements and an inspection of Eq.~\eqref{all_twist_non_amy} will reveal 
that this yields simply the pure vacuum contribution of Eq.~\eqref{O_00_3}.  
The second case is where we set $r,s$ the cross scattering to 0. 
Beyond all other trivial changes, this modifies the phase factor to yield, 
\bea
\Gamma &=& \lt[ \h(\zeta_I^- )  \lt\{ 1  
-  e^{ - ip^+ x_L \zeta_I^-  }  \rt\} \rt.  \label{r_0_s_0_phase} \\
&-&  \lt. \h( \zeta_I^- - y_I^- )  e^{ - ip^+ x_L y_I^-  }
-  \h( y_I^- - \zeta_I^- )  e^{ -ip^+ x_L \zeta_I^-   }  \rt] \nn \\
\ata \lt[ \h( \zeta_I^- )  \lt\{  1   
-  e^{  i p^+ x_L  \zeta_I^-  }  \rt\}  \rt. \nn\\
&-& \lt. \h( \zeta_I^- - y_I^- )  e^{ i p^+ x_L  y_I^- } 
- \h( y_I^- - \zeta_I^- )  e^{ i p^+ x_L \zeta_I^- }  \rt]. \nn 
\eea

We ignore the contributions that contain $y_{I}^{-}$  as there is 
no corresponding $p_{\perp}$  that appears in the amplitude, thus 
these will vanish when integrated over ${l_{q}}_{\perp}$. Also all 
terms which contain the phase factor $\exp{(\pm i p^{+} x_{L} \zeta_{I}^{-})}$ 
give leading contributions when $y_{I}^{-} > \zeta_{I}^{-}$. This allows 
the integrals over quark scatterings to be lumped together and included 
in the broadening of the two dimensional $\kd$-function over ${l_{q}}_{\perp}$. 
Thus the leading contributions (i.e., the least suppressed in $Q^{2}$) which 
contain at most one location $\zeta_{I}$ are given as,
\bea
&& \lt( 1 - e^{ -i p^{+} x_{L} \zeta_{I}^{-} } - \h (y_{I}^{-}  -  \zeta_{I}^{-}) e^{ -i p^{+} x_{L} \zeta_{I}^{-} } \rt) \nn \\
\ata  \lt( 1 - e^{ i p^{+} x_{L} \zeta_{I}^{-} } - \h (y_{I}^{-}  -  \zeta_{I}^{-}) e^{ i p^{+} x_{L} \zeta_{I}^{-} } \rt) \nn \\
&=&  2 + 3\h (y_I^- - \zeta_I^-) - 2 \cos ( p^+ x_L \zeta_I^- ) \nn \\
&-& 2 \h (y_I^- - \zeta_I^-)\cos ( p^+ x_L \zeta_I^- ) . \label{zeta_I_terms}
\eea
Along with these one will have to consider the terms which depend on the location 
$y_I^-$, i.e.,
\bea
\h( \zeta_I^- - y_I^- ) \lt[ - 2 \cos ( p^+ x_L y_I^- )  + 1 \rt]. \label{y_I_terms}
\eea

There is one extra factor of $\h(y_I^- - \zeta_I^-)$ without a cosine 
in Eq.~\eqref{zeta_I_terms} and 
one less factor if $\h( \zeta_I^-  - y_I^- )$ (without a cosine) in Eq.~\eqref{y_I_terms}. Since 
these terms have no extra dependence on any position they are essentially equal, i.e., 
these are symmetric across the line $y_I^- = \zeta_i^-$.  As a result,  
the term containing the extra factor or $\h(y_I^- - \zeta_I^-)$ can be converted to 
a term containing an $\h(y_I^- - \zeta_I^-)$ and included with Eq.~\eqref{y_I_terms}.
Such a conversion is required in the $l_\perp \gg k^i_\perp$  limit, as any term which 
cannot be cast in the final form, 
\bea
\int d l_\perp^2  \frac{ 2 - 2 \cos( x_L p^+ \zeta^- ) }{ l_\perp^4}, \label{LPM_finite}
\eea 
up to multiplicative factors, diverges as $1/l_\perp^4$ at small $l_\perp$.
Such a divergence calls into question the approximation of Taylor expanding in the 
softer momenta $k^i_\perp$.
Such a term also dominates over the vacuum radiation term at small $l_\perp$ and 
thus will invalidate the use of an in-medium DGLAP evolution equation 
to incorporate the effect of multiple radiations. All terms, at this order of 
approximation, are finite at $l_\perp \ra 0$ due to the destructive interference 
of the LPM effect.

We now consolidate terms which 
contain one more integral over light-cone position $\zeta_{I}^{-}$ 
than transverse derivative on the $\kd$ function, in 
Eq.~\eqref{all_twist_non_amy_2}. These are of two types, those which remain as $n \ra 0$ and those 
which vanish in this limit. Terms of the first type may be expressed as, 
\bea
\mbox{}\!\!\!\!\!\!&& \int_0^{L^-} d \zeta_I 4 D_G \lt[ 2 - 2\cos (p^+ x_L \zeta_I^- )  \rt]\\
\mbox{}\!\!\!\!\!\!\ata \!\!\!\!\!\!\!\!\sum_{m=1,n=0}^{\infty} \!\!\!\!\!\!
\frac{ \lt( \int_{\zeta_I^-}^{L^-} d \zeta^- D_G \nabla_{{l_q}_\perp}^2 \rt)^{m-1}  
\!\!\!\lt( \int_{0}^{L^-} d y^- D_Q \nabla_{{l_q}_\perp}^2 \rt)^{n}  \kd^2( {l_q}_\perp  ) }{(m-1)!n!} .\nn
\eea
Summing over $m,n$ reduces the second line of the above equation to the broadened Gaussian,
\bea
\frac{  \exp{\lt[  -\frac{ | {l_q}_\perp|^2 }{ 4 \int_{0}^{L^-} d y^- D_Q + 4 \int_{\zeta_I^-}^{L^-} d \zeta^- D_G }   \rt] }  }
{4 \pi \lt(   \int_{0}^{L^-} d y^- D_Q +  \int_{\zeta_I^-}^{L^-} d \zeta^- D_G   \rt) }.
\eea

The terms in Eq.~\eqref{all_twist_non_amy_2} which only exist for $n \geq 1$ and contain one more integral 
over the light-cone location $\zeta_{I}^{-}$  than transverse derivative on the $\kd$ function can be expressed as,   
\bea
\mbox{}\!\!\!\!\!\!&& \int_0^{L^-} d \zeta_I  4 D_G \lt[ 2 - 2\cos (p^+ x_L \zeta_I^- )  \rt]\\
\mbox{}\!\!\!\!\!\!\ata \!\!\!\!\!\!\!\!\sum_{m=1,n=1}^{\infty} \!\!\!\!\!\!
\frac{ \lt( \int_{\zeta_I^-}^{L^-} d \zeta^- D_G \nabla_{{l_q}_\perp}^2 \rt)^{m-1}  
\!\!\!\lt( \int_{\zeta_I^-}^{L^-} d y^- D_Q \nabla_{{l_q}_\perp}^2 \rt)^{n}  \kd^2( {l_q}_\perp  ) }{(m-1)!n!} .\nn
\eea
To sum over $m,n$ one of the $n$ integrals have to be extracted from the sum. Denoting this 
location as $y_L^-$, the result of the 
sum over $m,n$ for the second line of the above equation yields, 
\bea
\frac{  \int d y_L^-   D_Q \nabla_{{l_q}_\perp}^2  \exp{\lt[  -\frac{ | {l_q}_\perp|^2 }{ 4 \int_{\zeta_I^-}^{y_L^-} d y^- D_Q + 4 \int_{\zeta_I^-}^{L^-} d \zeta^- D_G }   \rt] }  }
{4 \pi \lt(   \int_{\zeta_I^-}^{y_L^-} d y^- D_Q +  \int_{\zeta_I^-}^{L^-} d \zeta^- D_G   \rt) } .
\eea
Integrating the above term over ${l_q}_\perp$ yields zero due to Eq.~\eqref{gaussian_rules}.

Combining these contributions, yields the leading medium dependent contribution to 
the gluon emission spectrum from multiple scattering in the small $y$ limit as, 
\bea
\frac{dN_1}{dy dl_{\perp}^{2}} &=& \frac{\A_{s} C_{F}}{2 \pi l_{\perp}^{4}} P(y)   \label{final_res} \\
\ata \int_{0}^{L^{-}} d \zeta^{-} 
4 D_{G}(\zeta^{-} ) \lt[  2  - 2\cos \lt( p^{+}  x_{L}  \zeta^{-}  \rt) \rt].\nn
\eea
Note that this is identical to the contribution in the previous subsection where we 
insisted that only the gluon scatters. 

The next-to-leading contribution  in the small $y$ limit may be derived similarly from the contributions 
in Eq.~\eqref{final_res_1}, including in addition the contributions which do not contain an exponential or a cosine factor, as 
\bea
\frac{dN_2}{dy dl_{\perp}^{2}} &=& -\frac{\A_{s} C_{F}}{2 \pi l_{\perp}^{4}} y P(y)   \label{final_res_y} \\
\ata \int_{0}^{L^{-}} d \zeta^{-} 
4 D_{GQ} (\zeta^{-} ) \lt[  2  - 2\cos \lt( p^{+}  x_{L}  \zeta^{-}  \rt) \rt].\nn
\eea
We now add the contributions where the initial quark scatters independently, i.e., where none of these 
scatterings interfere with scatterings on the radiated gluon or on the final quark. These diagrams 
contribute terms suppressed by one extra power of $l_{\perp}^{2}$ only in the limit where $r,s = 0$. 
These may be written down in analogy with Eq.~\eqref{final_res} by replacing $k_{\perp}^{i} \ra y p_{\perp}^{i}$
and $D_{G}$ with $D_{Q}$. As a result, this $y$ suppressed contribution is given as, 
\bea
\frac{dN_3}{dy dl_{\perp}^{2}} &=& \frac{\A_{s} C_{F}}{2 \pi l_{\perp}^{4}} y^{2}P(y)   \label{final_res_3} \\
\ata \int_{0}^{L^{-}} d \zeta^{-} 
4 D_{Q}(\zeta^{-} ) \lt[  2  - 2\cos \lt( p^{+}  x_{L}  \zeta^{-}  \rt) \rt].\nn
\eea

While the expressions above may be used in an energy loss calculation to determine the medium modified 
fragmentation function, it cannot be use for the Monte-Carlo simulation of jet propagation in medium. That 
will require the full double differential expressions in $l_{\perp}$ and $l_{q_{\perp}}$.  While these are 
not presented here, they may be straightforwardly derived from Eq.~\eqref{all_twist_non_amy} and 
will be discussed elsewhere.

Though not discussed in this article, the calculation of the medium modified fragmentation function or 
in the Monte-Carlo simulation of a hard jet, one requires probability decreasing contributions which 
counter the effect of the probability increasing expressions such as Eq.~\eqref{final_res}. In the case of
the medium modified fragmentation functions, these are essentially introduced as unitarity (or virtual) corrections.
In the case of jet simulations, one will traditionally derive a Sudakov factor~\cite{Armesto:2007dt} from the 
real splitting correction.


\section{Discussions, Conclusions and Outlook}


In this article, we have extended the higher twist formalism by deriving the
leading length-enhanced contribution to the single gluon splitting kernel at all-twist. 
Physically this includes the process of gluon emission from a hard 
quark which undergoes multiple scattering within the formation time 
of the produced collinear radiation.

We considered the propagation of a hard quark produced in DIS on a large 
nucleus. The hard virtual photon strikes a hard incoming quark in a nucleon which then 
proceeds to burrow through the nucleus in the direction of the hard photon. 
The hard quark is off-shell at the point of production and in the absence of 
the remaining nucleons would proceed to radiate multiple gluons prior to 
hadronization. We focused on one such hard emission $l_{\perp} \gg \Lambda_{QCD}$ 
which occurred inside the nucleus. The presence of the medium modifies the 
cross section to radiate the hard gluon due to multiple scattering of the propagating 
quark and the radiated gluon.

The main focus in the current effort has been on the calculation of 
the triple differential cross section to produced a single gluon with 
transverse momentum $l_{\perp}$, momentum fraction $y$ and a final outgoing quark with transverse 
momentum $l_{q_{\perp}}$ (and momentum fraction $1-y$). 
The series of diagrams which contain the leading 
length enhanced corrections to this process were identified. 
These diagrams were then evaluated in the collinear and hard 
approximation $q^- \ggg l_\perp \gg k_\perp$ where the first 
inequality suggests that the jet energy is logarithmically larger 
than the transverse momentum of the radiation, which by the second 
inequality is power enhanced compared to the transverse momentum 
imparted from the medium. 

Incorporating the various simplifications that result from the limits 
mentioned above, we evaluated the leading terms in the triple 
differential cross section. 
Besides the effect of multiple scattering on vacuum like splitting
diagrams, we also obtained new contributions which represent
the physical effects of a hard incoming gluon hitting a near on shell 
quark and the forcing it off shell and inducing a radiation. This 
interferes with the vacuum like emission as well as with 
other processes where on-shell quark splits, 
radiating a gluon and either the out-going quark or gluon goes off-shell.
The off-shell parton then quickly loses its excess virtuality by 
encountering another collision with an incoming hard gluon. 

After a discussion of length enhancements and color factors of 
different types of diagrams, we focused on diagrams which 
yield the leading power corrections to the single gluon cross section.
Among these, the leading corrections in the gluon momentum fraction $y$ 
with the largest color factors 
arose from processes where the final produced quark or gluon 
scattered independently or where there was interference between the 
scattering on the quark prior to emission and that on the quark or gluon 
after emission.

Following this, the double differential cross 
section to radiate a hard gluon was evaluated. This was evaluated in two 
separate cases: the case of only gluon scattering and the case which included 
all types of scattering. The leading corrections in a power expansion
involving the hard scale were then isolated. 
In these diagrams, the insistence that the transverse momentum of the radiated gluon 
is large in comparison with the transverse momentum of the incoming gluons 
from the medium make this calculation somewhat orthogonal to that of AMY.
Integrating the transverse momentum 
of the final produced quark reduced these contributions to those obtained 
from a single scattering analysis. This implies that the leading effect of 
multiple scattering is to broaden the transverse momentum distributions 
of the produced quark and gluon leaving the energy loss kernel unchanged. 
This justifies a similar assumption make in Ref.~\cite{Majumder:2006wi}, 
where it was assumed that the broadening of a radiated gluon could be considered 
independently from the emission cross section.

In this paper we have restricted the calculation to only the analytical 
component. The numerical evaluation of the medium modified 
fragmentation function followed by the phenomenological implementation 
in DIS on a large nucleus and in jets in high energy heavy ion collisions 
will be carried out in a separate  calculation. These will 
include the effect of multiple such stimulated emissions.

Along with the various terms outlined in the preceding section, 
we will also include a series of other terms which were not 
explicitly evaluated in this effort. Principle among these are the 
contributions suppressed by $Q^4$ (or rather $l_{\perp}^{4}$) and greater. 
The neglect of such 
contributions is justified at large $Q^2$. For a $Q^2 \sim 3 -5$GeV$^2$, 
as is the case for the HERMES experiments, and for large enough nuclei,
such power corrections may also have to be included.

The  DGLAP equation in vacuum, is a coupled equation which connects the 
evolution of the quark fragmentation function with that of the gluon. This will 
also be the case for energy loss in a medium. Thus a complete in-medium 
evolution will also require a calculation of the in-medium splitting function 
for a gluon to two gluons or a quark anti-quark pair. The derivation of 
such a splitting function will proceed in the same manner as in this 
article and will be presented in a future effort.

\section{Acknowledgments}

The authors thanks C. ~Gale, A.~Idilbi, S.~Jeon, Y.~Kovchegov, G.~Moore, B.~M\"{u}ller and G.-Y.~Qin  for helpful discussions. 
The author thanks U.~Heinz for a critical reading of a first draft of this manuscript and for discussions.
This work was supported in part by the U.S. Department of Energy under grant no. DE-FG02-01ER41190. 
This calculation was started while the author was employed at Duke University where it was supported in 
part by the U.S. Department of Energy under grant no. DE-FG02-05ER41367.


\section{Appendix A: $\lambda$ scaling of the $A^+$ field}


We use the linear response formula to ascertain the power counting of the $A^+$ field.
Suppressing the color superscripts we obtain, 
\bea
A^\mu (x) &=& \int d^4 y \mathcal{D}^{\mu \nu} (x-y) J_\nu (y) . \label{linear_resp}
\eea
In the equation above, $\mathcal{D}$ is the gluon propagator and at leading
order in the covariant gauge is given as, 
\bea
\mathcal{D}^{\mu \nu} (x -y ) &=& \int \frac{d^4 k}{ (2 \pi)^4} \frac{ - i \gmn e^{-i k \x ( x - y )}  }{ k^2 + i \e}. 
\label{covariant_gauge_gluon}
\eea
In Eq.~\eqref{linear_resp}, $J^{\nu} (y) = \psibar (y) \g^\nu \psi (y)$ is the current of partons in the target which generates the gluon field. 
The fermionic operator may be decomposed as,
\bea
\mbx\!\!\!\psi (y) \!\!&=&\!\! \int \frac{d p^+ d^2 p_\perp }{(2\pi)^3 \sqrt{p^+ + \frac{p_\perp^2}{2p^+}  }} 
\sum_s u^s(p) a_p^s e^{-ip\x y} + \ldots
\eea
The scaling of the fermionic operator depends on the range of momentum which are selected 
from the in-state by the annihilation operator. Note that this influences both the scaling of the 
annihilation operator $a_p$ as well as the bispinor $u(p)$. The power counting of the annihilation 
operator may be surmised from the standard anti-commutation relation, 
\bea
\{ a_p^r , {a_{p^\p}^s}^{\dag} \} = (2 \pi)^3 \kd^3 (\vp - \vp^\p) \kd^{rs}.
\eea
and the power counting of the bispinor from the normalization condition,
\bea
\sum_s u_p^s \bar{u}_p^s = \f p = \g^- p^+ + \g^+ p^- - \g_\perp \x p_\perp.
\eea 
Substituting the equation for the current in Eq.~\ref{covariant_gauge_gluon}, and integrating out 
$y$, we obtain, 

\bea
\mbx\!\!\!\!\!\!\!A^+ \!\!\!&\simeq&\! \!\!\!\int\!\! \frac{d^3 p d^3 q}{(2\pi)^6  \sqrt{p^+} \sqrt{q^+}}  
\frac{-i e^{-i(p-q)\x x} }{ (p-q)^2} a_q^\dag a_p 
\bar{u}(q) \g^+ u(q).
\eea
If the incoming and out going momenta $p$ and $q$ scale as collinear momenta in the ($+$)-direction, 
i.e., $p \sim Q(1,\lambda^2, \lambda)$, then we get, $\kd^3 ( \vp - \vp^\p ) \sim [ \lambda^2 Q^3 ]^{-1}$, as one 
of the momenta will involve the large ($+$)-component and the remaining are the small transverse 
components. Thus the annihilation (and creation) operator scales as $\lambda^{-1} Q^{-3/2}$. Also in the spin 
sum $\f \!\!p \sim Q$ and thus $u(p) \sim u(q) \sim Q^{1/2}$; one can check that the $\g^+$ 
projects out the large components in $u$ and  $\bar{u}$ in the expression $\bar{u}(q) \g^+ u(p)$. 
We also institute the Glauber condition that 
$p^+ - q^+ \sim \ld^2 Q$, $p^- - q^- \sim \ld^2 Q$ and $p_\perp - q_\perp \sim \ld Q$.

Using these scaling relations we correctly find that the bispinor scales as 
$\lambda Q^{3/2}$. However, to obtain the correct scaling of the gauge field $A^+$ 
one needs to institute the approximation that $q^+ = p^+ + k^+$ where $k^+ \sim \lambda^2 Q$.
This condition is introduced by insisting that the $(+)$ momentum of the incoming and outgoing 
state, which control the scaling of $a_q^\dag$ and $a_p$, are separated by $k^+ \sim \ld^2 Q$. This is used to 
shift the $dq^+ \ra d k^+$ and as a result we obtain the scaling of the $A^+$ field as $\ld^2 Q$.
Following a similar derivation, with the replacement $\g^+ \ra \g^\perp$ we obtain the scaling of the 
$A^\perp \sim \ld^3 Q$.


\begin{thebibliography}{99}
\bibitem{f2} 
A more correct notation would be to refer 
to this state as $| A- 1\rc$, ignoring the $1$ is appropriate for large $A$. These issues are dealt with more rigor in Sect.~V.
\end{thebibliography}
\end{document}